\documentclass[journal,12pt,onecolumn,draftclsnofoot,]{IEEEtran}

\usepackage[utf8]{inputenc}
\usepackage{amssymb}
\usepackage{amsmath,bm}
\usepackage{mathrsfs}
\usepackage{mathrsfs}
\usepackage{xcolor,xspace}
\usepackage{setspace}
\usepackage{scalerel}
\usepackage{theorem}
\usepackage{cite}
\usepackage{url}
\usepackage{orcidlink}
\usepackage{array,multirow}
\usepackage{pgfplots}

\usepackage[ruled,vlined,titlenumbered,linesnumbered]{algorithm2e}
\usepackage[nolist]{acronym}

\usepackage{makecell}
\usepackage{mleftright}
\usepackage{subcaption}

\usepackage{comment}

\usepackage{mathtools}
\mathtoolsset{showonlyrefs}
\usepackage{xcolor}

\usetikzlibrary{plotmarks,patterns}
\usetikzlibrary{positioning, arrows.meta, calc}

\definecolor{darkgreen}{RGB}{0,100,0}
\definecolor{darkblue}{RGB}{0,0,139}
\definecolor{darkolive}{RGB}{85,107,47}
\definecolor{darkpurple}{RGB}{128,0,128}
\definecolor{navyblue}{RGB}{0,0,128}
\definecolor{forestgreen}{RGB}{34,139,34}
\definecolor{mediumorchid}{RGB}{186,85,211}
\definecolor{crimson}{RGB}{220,20,60}
\definecolor{lightgray}{RGB}{211,211,211}

\pgfplotsset{
compat=1.17,
mystyle/.style={
    scale only axis,
    width=0.7\columnwidth,
    height=0.5\columnwidth,
    label style={inner sep=0, font=\normalsize}, 
    tick label style={font=\scriptsize},
    legend style={font=\scriptsize},
    mark size=3,
    major grid style={dashed},
    line width=0.8pt,
    axis line style = thin}
}

\tikzset{
errorblock/.style={
    pattern=north west lines,
    pattern color=black!50,
    draw=black!50,
    line width=0.5pt
},
nonerrorblock/.style={
    fill=white,
    draw=black!50,
    line width=0.5pt
}
}

\pgfplotsset{
  line1/.style={color=blue, dashed, line width=1pt},
line2/.style={color=red, dashdotted, line width=1pt, opacity=0.7},
line3/.style={color=olive, dotted, line width=1.2pt, dash pattern=on 2pt off 2pt},
  line4/.style={color=black, solid, line width=1.2pt},
}

\tikzset{
  simulated/.style={color=blue!50!cyan, thick, only marks, mark=*, mark size=1.5pt, mark options={solid, fill=blue!50!cyan}},
  lowerbound/.style={color=mediumorchid, dashed, thick},
  upperbound/.style={color=mediumorchid, dashdotted, thick},
  approximation/.style={color=orange!80!red!60!white, dotted, very thick},
}

\newcolumntype{M}[1]{>{\centering\arraybackslash}m{#1}}

\newtheorem{theorem}{Theorem}
\newtheorem{definition}{Definition}
\newtheorem{lemma}{Lemma}
\newtheorem{corollary}{Corollary}

\newtheorem{remark}{Remark}

\newtheorem{example}{Example}

\newcommand{\Fqm}{\ensuremath{\mathbb F_{q^m}}}

\newcommand{\Fq}{\ensuremath{\mathbb F_{q}}}

\newcommand{\F}{\ensuremath{\mathbb F}}
\newcommand{\ZZ}{\ensuremath{\mathbb{Z}}}
\newcommand{\NN}{\ensuremath{\mathbb{N}}}

\newcommand{\set}[1]{\ensuremath{\mathcal{#1}}}

\newcommand{\dmin}{\ensuremath{d}}

\newcommand{\dminsr}{\ensuremath{d}}

\SetKwComment{Comment}{}{}

\newcommand{\ext}{\ensuremath{\text{ext}}}

\newcommand{\defeq}{:=}

\newcommand{\inshots}{\ensuremath{\in\{1,\ldots,\shots\}}}
\newcommand{\inshotsarg}[1]{\ensuremath{\in\{1,\ldots,#1\}}}

\newcommand{\PLB}{\ensuremath{\mathrm{P}_{\mathrm{LB}}}}
\newcommand{\PUB}{\ensuremath{\mathrm{P}_{\mathrm{UB}}}}
\newcommand{\pFullRankB}{\ensuremath{p_{1}}}
\newcommand{\pFullRankBb}{\ensuremath{p_{2}}}

\DeclareMathOperator{\wt}{wt}
\DeclareMathOperator{\rk}{rk}
\DeclareMathOperator{\rkq}{rk_{q}}
\DeclareMathOperator{\rkqm}{rk_{q^m}}

\DeclareMathOperator{\supp}{supp}

\DeclareMathOperator{\diag}{diag}
\DeclareMathOperator{\REF}{REF}

\renewcommand{\vec}[1]{\ensuremath{\bm{#1}}}
\newcommand{\mat}[1]{\ensuremath{\bm{#1}}}

\renewcommand{\b}{\vec{b}}
\renewcommand{\c}{\vec{c}}

\newcommand{\g}{\vec{g}}

\newcommand{\n}{\vec{n}}

\renewcommand{\t}{\vec{t}}

\newcommand{\x}{\vec{x}}
\newcommand{\y}{\vec{y}}

\newcommand{\A}{\mat{A}}
\newcommand{\B}{\mat{B}}
\newcommand{\C}{\mat{C}}
\newcommand{\D}{\mat{D}}
\newcommand{\E}{\mat{E}}
\newcommand{\G}{\mat{G}}
\newcommand{\I}{\mat{I}}

\renewcommand{\H}{\mat{H}}
\newcommand{\M}{\mat{M}}

\renewcommand{\P}{\mat{P}}
\newcommand{\Q}{\mat{Q}}

\renewcommand{\S}{\mat{S}}

\newcommand{\X}{\mat{X}}
\newcommand{\Y}{\mat{Y}}

\newcommand{\0}{\ensuremath{\mathbf 0}}

\newcommand{\mycode}[1]{\ensuremath{\mathcal{#1}}}

\newcommand{\errorCode}{\mycode{E}}
\newcommand{\CCode}{\mycode{C}}
\newcommand{\CECode}{{{\mycode{S}}}}

\newcommand{\SumRankWeight}[1]{\ensuremath{\wt_{\Sigma R}^{(#1)}}}

\newcommand{\SumRankDist}[1]{d_{\ensuremath{\Sigma}R}^{(#1)}}

\newcommand{\RankSupp}{\supp_{R}}
\newcommand{\SumRankSupp}{\supp_{\Sigma R}}
\newcommand{\HammingSupp}{\supp_{H}}

\newcommand{\sep}{\ensuremath{\mid}}

\newcommand{\myspace}[1]{\mathcal{#1}}
\newcommand{\Rowspace}[1]{\ensuremath{\myspace{R}_{q}\!\left(#1\right)}}
\newcommand{\RowspaceFqm}[1]{\ensuremath{\myspace{R}_{q^m}\!\left(#1\right)}}

\newcommand{\quadbinom}[2]{\left[\genfrac{}{}{0pt}{}{#1\vphantom{N_N}}{#2\vphantom{N}}\right]_{q}}

\newcommand{\oh}[1]{\bnd{O}{#1}}

\newcommand{\bnd}[2]{\ensuremath{#1\mathopen{}\left(#2\right)\mathclose{}}}

\newcommand{\wdecomp}[1]{\ensuremath{\mathcal{T}_{#1}}}

\newcommand{\shot}[2]{\ensuremath{{#1}^{(#2)}}}

\newcommand{\intOrder}{\ensuremath{s}}

\newcommand{\shots}{\ensuremath{\ell}}
\newcommand{\shotLength}{\ensuremath{\eta}}

\newcommand{\MeKap}{Metzner--Kapturowski}

\DeclareMathOperator{\mksum}{\ensuremath{\bigoplus}}

\DeclareMathOperator{\NM}{NM}

\newcommand{\T}{\vec{T}}

\newcommand{\Hsub}{\ensuremath{\Hce}}
\newcommand{\Hce}{\ensuremath{\H_{\CECode}}}

\newcommand{\errWeight}{\ensuremath{t}}
\newcommand{\errWeightVec}{\ensuremath{\t}}

\begin{acronym}
 \acro{BCH}{Bose--Chaudhuri--Hocquenghem}
 \acro{BMD}{bounded minimum distance}
 \acro{ESP}{error span polynomial}
 \acro{ELP}{error locator polynomial} 
 \acro{SRS}{skew Reed--Solomon}
 \acro{ISRS}{interleaved skew Reed--Solomon}
 \acro{lclm}{least common left multiple}
 \acro{LRS}{linearized Reed--Solomon}
 \acro{LLRS}{lifted linearized Reed--Solomon}
 \acro{ILRS}{interleaved linearized Reed--Solomon}
 \acro{LILRS}{lifted interleaved linearized Reed--Solomon}
 \acro{MDS}{maximum distance separable}
 \acro{MSRD}{maximum sum-rank distance}
 \acro{MSD}{maximum skew distance}
 \acro{RS}{Reed--Solomon}
 \acro{REF}{row echelon form}
 \acro{LEEA}{linearized extended Euclidean algorithm}
 \acro{SEEA}{skew extended Euclidean algorithm}
 \acro{VSD}{vector-symbol decoding}
 \acro{AG}{algebraic-geometry}
 \acro{KEM}{key encapsulation mechanism}
 \acro{ISD}{information-set-decoding}
 \acro{NIST}{National Institute of Standards and Technology}
\end{acronym}

\title{An Error-Code Perspective on Metzner--Kapturowski-like Decoders}

\author{
 Thomas Jerkovits~\orcidlink{0000-0002-7538-7639},~\IEEEmembership{Graduate Student Member,~IEEE, }%
 Felicitas Hörmann~\orcidlink{0000-0003-2217-9753},~\IEEEmembership{Graduate Student Member,~IEEE, }%
 Hannes Bartz~\orcidlink{0000-0001-7767-1513},~\IEEEmembership{Member,~IEEE}%
}

\begin{document}

\maketitle

\begin{abstract}
In this paper we consider a Metzner–Kapturowski-like decoding algorithm for high-order interleaved sum-rank-metric codes, offering a novel perspective on the decoding process through the concept of an error code. The error code, defined as the linear code spanned by the vectors forming the error matrix, provides a more intuitive understanding of the decoder's functionality and new insights.

The proposed algorithm can correct errors of sum-rank weight up to $d-2$, where $d$ is the minimum distance of the constituent code, given a sufficiently large interleaving order. The decoder's versatility is highlighted by its applicability to any linear constituent code, including unstructured or random codes. The computational complexity is $\oh{\max\{n^3, n^2s\}}$ operations over $\Fqm$, where $n$ is the code length and $s$ is the interleaving order.

We further explore the success probability of the decoder for random errors, providing an efficient algorithm to compute an upper bound on this probability. Additionally, we derive bounds and approximations for the success probability when the error weight exceeds the unique decoding radius, showing that the decoder maintains a high success probability in this regime.

Our findings suggest that this decoder could be a valuable tool for the design and security analysis of code-based cryptosystems using interleaved sum-rank-metric codes. The new insights into the decoding process and the high success probability of the algorithm even beyond the unique decoding radius underscore its potential to contribute to various coding-related applications.

\end{abstract}


\section{Introduction}
\label{sec:introduction}

\IEEEPARstart{T}{he} need for post-quantum cryptography has become increasingly important due to recent advances in the design and realization of quantum computers.
This has led to the \ac{NIST}'s post-quantum cryptography standardization process, which has shown that many promising candidates for \acp{KEM} belong to the family of code-based systems.
Three of these candidates are still in the current 4th round~\cite{NISTreport2022}.

Most code-based cryptosystems are based on the McEliece cryptosystem~\cite{McEliece1978}, which uses a public code that can only be efficiently decoded with knowledge of the secret key as its trapdoor.
However, a major drawback of code-based cryptosystems is their large public key sizes when compared to other schemes based on, e.g., lattices or isogenies.
Completely unstructured (i.e., random) codes require a large key size,
while the usage of highly structured codes often results in vulnerabilities that can be exploited in structural attacks.

Interleaving has been proposed as one approach to mitigate the key-size issue in variants of the McEliece cryptosystem based on interleaved codes in the Hamming and rank metric~\cite{elleuch2018public,holzbaur2019decoding,renner2019interleavingloidreau}.
By allowing for a larger decoding radius and a higher error weight, denoted by $\errWeight$, interleaving increases the attack cost for the same-sized parameters. This effectively reduces the public-key size while maintaining the same security level.
The interleaving order, denoted by $\intOrder$, plays a crucial role in the decoding process and the overall performance of the cryptosystem.

There exist list and probabilistic unique decoders for interleaved \ac{RS} codes in the Hamming metric~\cite{krachkovsky1997decoding}, for interleaved Gabidulin codes in the rank metric~\cite{Loidreau_Overbeck_Interleaved_2006}, and for interleaved \ac{LRS} codes in the sum-rank metric~\cite{bartz2022fast}.
However, these decoders are tailored to a particular code family and explicitly exploit the code structure.
In contrast, the \MeKap~decoder, originally proposed in the Hamming metric, exploits a high interleaving order $\intOrder$ to successfully decode errors with high probability, independent of the constituent code~\cite{metzner1990general}.
This purely linear-algebraic decoder has been further studied and generalized to the rank~\cite{renner2021decoding} and sum-rank metric~\cite{jerkovits2023highorder}.

The sum-rank metric is part of a metric family that includes both the Hamming and the rank metric as special cases and can be seen as a blend of these two metrics.
In this framework, codeword vectors are discretely organized into blocks of equal length.
The sum-rank metric offers a balanced approach between the Hamming and rank metric, potentially making it more challenging for adversaries to exploit system vulnerabilities.

Compared to the Hamming metric, the rank metric has a higher generic decoding complexity for a given error weight. However, for the same decoding ``attack complexity'', it allows for smaller error weights and therefore smaller code parameters, which in turn leads to smaller key sizes.

However, many rank-metric cryptosystems rely on highly structured codes, which have been subject to attacks and have been broken in some cases.
Significantly, many attacks effective in the Hamming metric may prove ineffective in the rank metric and vice versa~(e.g.~\cite{hormannDistinguishingRecoveringGeneralized2023a}).
Given this unique attribute, the sum-rank metric offers a balanced approach, potentially making it more resistant to attacks that exploit vulnerabilities specific to either the Hamming or rank metric.
By carefully choosing the block size and the number of blocks, the sum-rank metric can be tuned to achieve a desired balance between security and key size.
This flexibility makes the sum-rank metric an attractive option for designing code-based cryptosystems that are secure against quantum and classical attacks while maintaining practical key sizes.

The goal of this paper is twofold: (1) to provide more intuition about Metzner--Kapturowski-like decoders by using an interpretation involving an \emph{error code} and (2) to extend the results from our previous work~\cite{jerkovits2023highorder}.
For (1), we make a connection to a code that we call the error code, which is spanned by the $\intOrder$ rows of the error matrix.
This perspective allows us to provide a more intuitive understanding of the decoding process by relating it to properties of the error code.
Furthermore, this new error-code perspective enables us to simplify proofs and derive new interpretations for the special cases in the Hamming and rank metric.
For (2), we investigate the success probability for high interleaving orders but randomly chosen errors that are not necessarily full-rank.
We provide an algorithm to efficiently compute and upper bound this probability and present a more precise analysis of the decoding condition and bounds on its occurrence probability for arbitrary error weight.
We derive lower and upper bounds, as well as an approximation for the success probability, in the case of $\errWeight \geq \dmin - 1$, where $\dmin$ is the minimum distance of the underlying code, using random coding techniques. We also provide simulation results to support the tightness of our analysis.
The outcome of this analysis reveals that the success probability remains relatively high even for $\errWeight \geq \dmin-1$.
We provide examples to illustrate these bounds and approximation.

We present a \MeKap-like decoding algorithm for high-order interleaved sum-rank-metric codes with an arbitrary linear constituent code that can correct errors of sum-rank weight $\errWeight$ up to $\errWeight < n-k$, where $n$ and $k$ denote the length and dimension of the linear constituent code, respectively.
Remarkably, the proposed algorithm works for any linear constituent code, including unstructured or random codes, making it highly versatile.
The computational complexity of the algorithm is in the order of $\oh{\max\{n^3, n^2\intOrder\}}$ operations over $\Fqm$.
Note that the decoding complexity is independent of the code structure of the constituent code since the proposed algorithm exploits properties of high-order interleaving only.
This gives valuable insights for the design of McEliece-like cryptosystems based on interleaved codes in the sum-rank metric.
Since the sum-rank metric generalizes both the Hamming and the rank metric, the original \MeKap~decoder~\cite{metzner1990general} as well as its rank-metric analog~\cite{renner2021decoding} can be recovered from our proposal.

\section{Preliminaries}

\subsection{Notation}

Let $q$ be a power of a prime and let $\Fq$ denote the finite field of order $q$ and $\Fqm$ an extension field of degree $m$.
We use $\Fq^{a\times b}$ to denote the set of all $a\times b$ matrices over $\Fq$ and $\Fqm^b$ for the set of all row vectors of length $b$ over $\Fqm$.

Let $\vec{b} = \left[b_1,\ldots,b_m\right] \in \Fqm^m$ be a fixed (ordered) basis of $\Fqm$ over $\Fq$.
We denote by $\ext(\alpha)$ the column-wise expansion of an element $\alpha \in \Fqm$ over $\Fq$ (with respect to $\b$), i.e.,
\begin{equation*}
    \ext : \Fqm \rightarrow \Fq^{m \times 1}
\end{equation*}
such that $\alpha = \vec{b} \cdot \ext(\alpha)$.

For a vector $\vec{v} = \left[v_1,\ldots,v_n\right] \in \Fqm^n$, the notation is extended element-wise as follows
\begin{equation*}
    \ext(\vec{v}) = \left[\ext(v_1),\ldots,\ext(v_n)\right] \in \Fq^{m \times n},
\end{equation*}
where $\ext(\vec{v})$ is a matrix with columns $\ext(v_i) \in \Fq^{m \times 1}$ for $i=1,\ldots,n$.

Similarly, for a matrix $\M = \left[M_{i,j}\right] \in \Fqm^{k \times n}$, the notation is extended element-wise as follows
\begin{equation*}
    \ext(\M) = \left[\begin{array}{ccc}
        \ext(M_{1,1}) & \cdots & \ext(M_{1,n}) \\
        \vdots & \ddots & \vdots \\
        \ext(M_{k,1}) & \cdots & \ext(M_{k,n})
    \end{array}\right] \in \Fq^{mk \times n},
\end{equation*}
where $\ext(\M)$ is a matrix obtained by replacing each element $M_{i,j}$ of $\M$ with its corresponding column-wise expansion $\ext(M_{i,j}) \in \Fq^{m \times 1}$, for $i=1,\ldots,k$ and $j=1,\ldots,n$.

For a matrix $\A$ of size $a \times b$ and entries $A_{i,j}$ for $i \in\{1,\ldots,a\}$  and $j \in\{1,\ldots,b\}$, we define the submatrix notation
\begin{equation*}
    \A_{[c:d],[e:f]}\defeq
    \begin{bmatrix}
        A_{c,e} & \dots & A_{c,f}
        \\
        \vdots & \ddots & \vdots
        \\
        A_{d,e} & \dots & A_{d,f}
    \end{bmatrix}.
\end{equation*}

The $\Fqm$-linear row space of a matrix $\A$ over $\Fqm$ is denoted by $\RowspaceFqm{\A}$.
Its $\Fq$-linear row space is defined as $\Rowspace{\A} \defeq \myspace{R}_q{(\ext{(\A)})}$.
We denote the row-echelon form of $\A$ as $\REF(\A)$ 

\subsection{Sum-Rank-Metric Codes}
Let $\n=[n_1,\dots, n_\shots]\in\NN^\shots$ with $n_i > 0$ for all $i \inshots$ be a length partition\footnote{Note that this is also known as (integer) composition into exactly $\shots$ parts in combinatorics.} of $n$, i.e., $n = \sum_{i=1}^{\shots} n_i$.
Further, let $\x=[\x^{(1)} \mid \x^{(2)} \mid \dots \mid \x^{(\shots)}]\in\Fqm^n$ be a vector over a finite field $\Fqm$ with $\x^{(i)}\in\Fqm^{n_i}$ for each $i \inshots$. 
The rank of each block $\x^{(i)}$ is defined as $\rkq(\x^{(i)}) \defeq \rkq(\ext(\x^{(i)}))$, where $\ext(\x^{(i)})\in\Fq^{m\times n_i}$ is the column-wise expansion of $\x^{(i)}$ over $\Fq$.

The \emph{sum-rank weight} of $\x$ with respect to the length partition $\n$ is defined as
\begin{equation}\label{eq:defsumrankweight}
    \SumRankWeight{\n}(\x) \defeq \sum_{i=1}^{\shots}\rk_q(\x^{(i)}),
\end{equation}
and the \emph{sum-rank distance} between two vectors $\x, \y \in \Fqm^n$ is given by
\begin{equation}
    \SumRankDist{\n}(\x,\y) \defeq \SumRankWeight{\n}(\x-\y).
\end{equation}
Note that the sum-rank metric coincides with the Hamming metric when $\shots=n$ (i.e., $n_i=1$ for all $i\inshots$) and reduces to the rank metric when $\shots=1$.

An \emph{$\Fqm$-linear sum-rank-metric code} $\mycode{C}$ is an $\Fqm$-subspace of $\Fqm^{n}$.
It has length $n$ (with respect to a length partition $\n$), dimension $k \defeq \dim_{q^m}(\mycode{C})$ and minimum
(sum-rank) distance 
\begin{equation}
\dminsr \defeq \min \{\SumRankDist{\n}(\x,\y) : \x, \y \in \mycode{C}, \x \neq \y\}.
\end{equation}
To emphasize its parameters, we write $\mycode{C}[\n, k, \dminsr]$ in the following.

\subsection{Interleaved Sum-Rank-Metric Codes and Channel Model}
A (vertically) $s$-interleaved code is a direct sum of $s$ codes of the same length $n$.
In this paper we consider \emph{homogeneous} interleaved codes, i.e., codes obtained by interleaving codewords of \emph{a single} constituent code.

\begin{definition}[Interleaved Sum-Rank-Metric Code]
    Let $\mycode{C}[\n,k,\dminsr]\subseteq\Fqm^n$ be an $\Fqm$-linear sum-rank-metric code of length $n$ with length partition $\n=\left[n_1,n_2,\dots,n_\shots\right]\in\NN^\shots$ and minimum sum-rank distance $\dminsr$.
    Then the corresponding (homogeneous) $\intOrder$-interleaved code is defined as
    \begin{equation*}
        \mycode{IC}[\intOrder;\n,k,\dminsr]
        \defeq
        \left\{
        \begin{bmatrix}
            \c_1
            \\[-4pt]
            \vdots
            \\[-4pt]
            \c_\intOrder
        \end{bmatrix}:
        \c_j\in \mycode{C}[\n,k,\dminsr]
        \right\}
        \subseteq\Fqm^{\intOrder \times n}.
    \end{equation*}
\end{definition}

Each codeword $\C\in\mycode{IC}[\intOrder;\n,k,\dminsr]$ can be written as
\begin{equation*}
    \mat{C}=
    \left[
    \begin{array}{c|c|c|c}
        \c_1^{(1)} & \c_1^{(2)} & \dots & \c_1^{(\shots)}
        \\
        \vdots & \vdots & \ddots & \vdots
        \\
        \c_\intOrder^{(1)} & \c_\intOrder^{(2)} & \dots & \c_\intOrder^{(\shots)}
    \end{array}
    \right]\in\Fqm^{\intOrder \times n}
\end{equation*}
or equivalently as
\begin{equation*}
    \mat{C}=
    \left[\vec{C}^{(1)} \mid \mat{C}^{(2)} \mid \dots \mid \mat{C}^{(\shots)}\right]
\end{equation*}
where
\begin{equation}
    \label{eq:comp_mat_ILRS}
    \vec{C}^{(i)}\defeq
    \begin{bmatrix}
        \c_1^{(i)}
        \\
        \c_2^{(i)}
        \\
        \vdots
        \\
        \c_\intOrder^{(i)}
    \end{bmatrix}
    \in\Fqm^{\intOrder\times n_i}
\end{equation}
for all $i\inshots$.

As a channel model we consider the additive sum-rank channel
\begin{equation}
    \Y = \C + \E
\end{equation}
where
\begin{equation}
    \E=\left[\E^{(1)} \, | \, \E^{(2)} \, | \, \dots \, | \, \E^{(\shots)} \right]\in\Fqm^{\intOrder \times n}
\end{equation}
with $\E^{(i)}\in\Fqm^{\intOrder \times n_i}$ and $\rk_q(\E^{(i)})=t_i$ for all $i\inshots$ is an error matrix with $\SumRankWeight{\n}(\E)=t\defeq\sum_{i=1}^{\shots}t_i$.

\subsection{The Error Support}
Let $\E\in\Fqm^{\intOrder \times n}$ be the error matrix with $\SumRankWeight{\n}(\E)=\errWeight$. Then $\E$ can be decomposed as
\begin{equation}\label{eq:error_decomp}
    \E = \A \B,
\end{equation}
where $\A = \left[\A^{(1)} \sep \A^{(2)} \sep \dots \sep \A^{(\shots)} \right]\in\Fqm^{\intOrder \times \errWeight}$ is a block matrix with submatrices $\A^{(i)}\in\Fqm^{\intOrder\times \errWeight_i}$ satisfying $\rkq(\A^{(i)})=\errWeight_i$, and
\begin{equation}\label{eq:def_B}
    \B = \diag{(\B^{(1)},\dots,\B^{(\shots)})} \in \Fq^{\errWeight \times n}
\end{equation}
is a block-diagonal matrix with submatrices $\B^{(i)}\in\Fq^{\errWeight_i \times n_i}$ satisfying $\rkq(\B^{(i)})=\errWeight_i$ for all $i\inshots$ (see~\cite[Lemma 10]{puchingerGenericDecodingSumRank2022}).

The rank support $\RankSupp\left(\E^{(i)}\right)$ and the dual rank support $\RankSupp^{\perp}{\left(\E^{(i)}\right)}$ of one block $\E^{(i)}$ for $i \inshots$ are defined as the row space of $\E^{(i)}$ and its orthogonal complement, respectively
\begin{alignat}{2}
    \RankSupp\left(\E^{(i)}\right) &\defeq \Rowspace{\E^{(i)}} = \Rowspace{\B^{(i)}}, \\
    \RankSupp^{\perp}{\left(\E^{(i)}\right)} &\defeq \Rowspace{\E^{(i)}}^{\perp} = \Rowspace{\B^{(i)}}^{\perp}.
\end{alignat}
The second equality in each line follows from~\eqref{eq:error_decomp} and~\cite[Theorem 1]{matsaglia_equalities_1974}.

The sum-rank support of the error $\E$ with sum-rank weight $\errWeight$ is then defined as
\begin{align}\label{eq:sumranksupport}
    \SumRankSupp(\E)
    \defeq
    &\RankSupp{\left(\E^{(1)}\right)} \times \RankSupp{\left(\E^{(2)}\right)} \times \dots \times \RankSupp{\left(\E^{(\shots)}\right)}
    \\
    =&\Rowspace{\B^{(1)}} \times \Rowspace{\B^{(2)}} \times \dots \times \Rowspace{\B^{(\shots)}}.
\end{align}
Additionally, we define the dual sum-rank support as
\begin{align}\label{eq:complsumranksupport}
    {\SumRankSupp^{\perp}{(\E)}} \defeq
    &\RankSupp^{\perp}{\left(\E^{(1)}\right)} \times \RankSupp^{\perp}{\left(\E^{(2)}\right)} \times \dots \times \RankSupp^{\perp}{\left(\E^{(\shots)}\right)}
    \\
    =&\Rowspace{\B^{(1)}}^{\perp} \times \Rowspace{\B^{(2)}}^{\perp} \times \dots \times \Rowspace{\B^{(\shots)}}^{\perp}.
\end{align}

Given two supports $\SumRankSupp(\E_1)$ and $\SumRankSupp(\E_2)$, we denote
\begin{equation}
\SumRankSupp(\E_1) \subseteq \SumRankSupp(\E_2)    
\end{equation}
if $\RankSupp{\left(\E_1^{(i)}\right)} \subseteq \RankSupp{\left(\E_2^{(i)}\right)}$ holds for all $i\inshots$. The notation $\subset$ follows the same principle but implies a strict subset.

Finally, we define 
\begin{equation}
    \Fq^{\n} \defeq \Fq^{n_1} \times \dots \times \Fq^{n_\shots}.
\end{equation}

\section{Decoding of High-Order Interleaved Sum-Rank-Metric Codes} \label{sec:decoder}

In this section, we propose a \MeKap-like decoder for the sum-rank metric, which generalizes the decoders presented in~\cite{metzner1990general, puchinger2019decoding, renner2021decoding}. The proposed decoder can correct errors of sum-rank weight $\errWeight$ up to $\dminsr-2$ in general.
Additionally, under specific conditions, the decoder can correct errors of sum-rank weight $\errWeight$ up to $n-k-1$, where $n$ is the length of the code and $k$ is the dimension of the code. The following assumptions are required for the decoder to succeed:

\begin{itemize}
\item \emph{High-order condition:} The interleaving order $\intOrder$ is greater than or equal to the sum-rank weight of the error, i.e., $\intOrder \geq \errWeight$.
\item \emph{Full-rank condition:} The error matrix has full $\Fqm$-rank, i.e., $\rkqm(\E) = \errWeight$.
\end{itemize}

It is worth noting that the full-rank condition automatically implies the high-order condition, as the $\Fqm$-rank of a matrix $\E \in \Fqm^{\intOrder \times n}$ cannot exceed the interleaving order $\intOrder$.

Throughout this section, we consider a homogeneous $\intOrder$-interleaved sum-rank-metric code $\mycode{IC}[\intOrder;\n,k,\dminsr]$ over $\Fqm$ with a constituent code $\CCode[\n,k,\dminsr]$ defined by a parity-check matrix 
\begin{equation}
    \H = \left[\H^{(1)}\sep\H^{(2)}\sep\ldots\sep\H^{(\shots)}\right]\in\Fqm^{(n-k)\times n}
\end{equation}
with $\H^{(i)}\in\Fqm^{(n-k) \times n_i}$.
The goal is to recover a codeword $\C\in\mycode{IC}[\intOrder;\n,k,\dminsr]$ from the matrix
\begin{equation*}
    \Y = \C + \E \in \Fqm^{\intOrder \times n}
\end{equation*}
that is corrupted by an error matrix $\E$ of sum-rank weight $\SumRankWeight{\n}(\E)=\errWeight$ assuming the \emph{high-order} and \emph{full-rank} conditions.

As with the original Metzner--Kapturowski algorithm and its adaptation to the rank metric, the presented decoding algorithm consists of two steps:
\begin{enumerate}
    \item The decoder determines the error support $\SumRankSupp(\E)$.
    \item Erasure decoding is performed using the syndrome matrix $\S = \H\Y^\top= \H \E^{\top}$ to recover the error $\E$ itself.
\end{enumerate}

The following result is adapted from~\cite{puchingerGenericDecodingSumRank2022} and shows how the error matrix $\E$ can be reconstructed from the sum-rank support $\SumRankSupp(\E)$ and the syndrome matrix $\S$. We relax the original condition to make the result more applicable.

\begin{lemma}[Column-Erasure Decoder~{\cite[Theorem 13]{puchingerGenericDecodingSumRank2022}}]\label{lem:col_erasure_decoding}
    Let $\B = \diag{(\B^{(1)},\dots,\B^{(\shots)})}\in\Fq^{\errWeight\times n}$ be a basis of the error support $\SumRankSupp(\E)$ of the error matrix $\E\in\Fqm^{\intOrder\times n}$, and let $\S = \H\E^\top \in \Fqm^{(n-k)\times \intOrder}$ be the corresponding syndrome matrix.
    
    Assume that $\H\B^\top$ is full-rank. Then, the error matrix $\E$ can be uniquely recovered as $\E = \A\B$, where $\A\in\Fqm^{\intOrder\times \errWeight}$ is the unique solution of the linear system
    \begin{equation}\label{eq:syndrome_system}
        \S = (\H\B^{\top})\A^\top.
    \end{equation}
    
    Furthermore, $\E$ can be computed in $\oh{(n-k)^3 m^2}$ operations over $\Fq$.
\end{lemma}

\begin{remark}
    From \cite[Lemma 12]{puchingerGenericDecodingSumRank2022}, it directly follows that for $\errWeight < \dminsr$, the condition that $\H\B^\top$ is full-rank is always satisfied.
\end{remark}

\subsection{Recovering the Error Support}

Let $\errWeightVec = [\errWeight_1,\ldots,\errWeight_\shots]$ denote the rank profile of the error matrix $\E$, where $\errWeight_i = \rkq(\E^{(i)})$ for $i\inshots$. In the following, we assume that $\E$ fulfills the full-rank condition, i.e., its $\Fqm$-rank is equal to its sum-rank weight $\errWeight$. Note that the full-rank condition is satisfied if and only if $\rkqm(\A)=\errWeight$ for an every $\A\in\Fqm^{\intOrder\times \errWeight}$ as in~\eqref{eq:error_decomp}.
Under these assumptions, we have that the rows of $\E$ span an $\Fqm$-linear $[\n,\errWeight]$ code, denoted as
\begin{equation}\label{eq:def_error_code}
\errorCode \defeq \RowspaceFqm{\E},
\end{equation}
which we refer to as the \emph{error code}.

Let $\G_{\errorCode}\in \Fqm^{t \times n}$ denote the generator matrix of $\errorCode$. Note that we can decompose $\G_{\errorCode}$ as
\begin{equation}\label{eq:gen_error_code}
\G_{\errorCode} = \A_{\errorCode} \B,
\end{equation}
where $\A_{\errorCode} = \left[\A_{\errorCode}^{(1)}\sep\ldots\sep\A_{\errorCode}^{(\shots)}\right]\in\Fqm^{\errWeight\times \errWeight}$ with $\rkqm(\A_\errorCode)=\errWeight$ and $\B$ is the same matrix as defined in the error decomposition~\eqref{eq:error_decomp} and~\eqref{eq:def_B}. Each block $\A_{\errorCode}^{(i)}$ is a matrix of size $\errWeight \times \errWeight_i$. The rank profile $\errWeightVec$ determines the ranks of the individual blocks $\A_{\errorCode}^{(i)}$, i.e., $\rkqm(\A_{\errorCode}^{(i)}) = \errWeight_i$.

It follows directly from the definition~\eqref{eq:def_error_code} of the error code, that 
\begin{equation}
    \SumRankSupp{(\errorCode)} = \SumRankSupp{(\E)}.
\end{equation}
Because of this property, we say that the error code $\errorCode$ is \emph{support-restricted by the row support of $\E$} with $\errorCode\subset \Fq^{\n}$.

Let us now consider the parity-check matrix of the error code $\errorCode$, denoted by $\H_\errorCode\in\Fqm^{(n-\errWeight)\times n}$.
By definition of the parity-check matrix, we have $\G_\errorCode\H_\errorCode^\top=\bm{0}$.
\begin{lemma}
    Let $\H_\errorCode=\left[\H_\errorCode^{(1)}\sep\ldots\sep\H_\errorCode^{(\shots)}\right]\in\Fqm^{(n-\errWeight)\times n}$ be the parity-check matrix of the $[\n, \errWeight]$ error code $\errorCode$ with length partition $\n$. Then, we have
    \begin{equation}
     \SumRankSupp(\H_\errorCode) = \SumRankSupp^{\perp}{(\E)}.
    \end{equation}
\end{lemma}
\begin{IEEEproof}
    Since $\H_\errorCode$ is a parity-check matrix of $\errorCode$, we have $\rkqm(\H_\errorCode) = n-\errWeight$.
    With respect to the sum-rank metric, we can partition the parity-check matrix of the error code as
    \begin{equation}\label{eq:Herror_code_split}
        \H_\errorCode = \left[\H_\errorCode^{(1)}\sep\ldots\sep\H_\errorCode^{(\shots)}\right]
    \end{equation}
    such that $\H_\errorCode^{(i)}\in\Fqm^{(n-\errWeight)\times n_i}$ for all $i\inshots$.
    
    To satisfy the check equations, we must have
    \begin{equation}\label{eq:proof_check_equations_sr}
        \G_\errorCode\H_\errorCode^\top = \bm{0} \;\Leftrightarrow\; (\A_\errorCode\B)\H_\errorCode^\top = \bm{0} \;\Leftrightarrow\; \B\H_\errorCode^\top = \bm{0}.
    \end{equation}
    From~\eqref{eq:Herror_code_split} and the block-diagonal structure of $\B$ (see~\eqref{eq:def_B}), it follows that
    \begin{equation}
        \B^{(i)} {\H_\errorCode^{(i)}}^\top = \bm{0} \quad\forall i\inshots.
    \end{equation}
    By the rank-nullity theorem and since $\B^{(i)}$ is over $\Fq$, we have
    $\dim\left(\Rowspace{\H_\errorCode^{(i)}}\right) \leq n_i - \errWeight_i$ for all $i\inshots$. However, since $\H_\errorCode$ must have $n-\errWeight$ many $\Fqm$-linearly independent rows and $\sum_{i=1}^{\shots} n_i - \errWeight_i = n - \errWeight$, we conclude that $\dim\left(\Rowspace{\H_\errorCode^{(i)}}\right) = n_i - \errWeight_i$, and hence
    \begin{equation}
        \Rowspace{\H_\errorCode^{(i)}} = \Rowspace{\B^{(i)}}^\top \quad \forall i\inshots.
    \end{equation}
    By the definition of the sum-rank support, this concludes the proof.
\end{IEEEproof}

\begin{theorem}\label{thm:errorsupportinFqHdual}
    Let $\CCode$ be an $\Fqm$-linear $[\n, k]$ sum-rank-metric code with generator matrix $\G\in\Fqm^{k\times n}$, parity-check matrix $\H\in\Fqm^{(n-k)\times n}$, and minimum sum-rank distance $\dminsr$. Let $\E=\A\B\in\Fqm^{\intOrder \times n}$ be a matrix with $\A\in\Fqm^{\intOrder\times \errWeight}$, $\B\in\Fq^{\errWeight\times n}$, $\rkqm(\E)=\errWeight$, and $\SumRankWeight{\n}{(\E)}=\errWeight$. Let $\errWeight\leq n-k-1$ and suppose that
    \begin{equation}\label{eq:rankBandb}
        \rkqm{\left(\H \left[\begin{array}{c}
        \B \\ \hline
        \b
        \end{array}\right]^\top\right)} = \errWeight + 1 \quad \forall\,\b \in \Fq^{\n} \setminus\SumRankSupp{(\E)} \;\;\text{s.t.}\;\SumRankWeight{\n}{(\b)} = 1.
    \end{equation}
    Further, denote by $\G_\errorCode\in\Fqm^{\errWeight\times n}$ the generator matrix of the error code $\errorCode \defeq \RowspaceFqm{\E}$. Consider the $\Fqm$-linear code $\CECode = \errorCode + \CCode$
    defined as
    \begin{equation}\label{eq:codeanderrorsumcode}
    \CECode \defeq \RowspaceFqm{\G_\CECode}
    \end{equation} 
    with generator matrix 
    \begin{equation}\label{eq:genCECode}
        \G_\CECode \defeq \left[\begin{array}{c}
    \G \\ \hline
    \E
    \end{array}\right].
    \end{equation}
    Then, for any valid parity-check matrix $\Hce\in\Fqm^{(n-k-\errWeight)\times n}$ of the $\Fqm$-linear $[\n, k+\errWeight]$ sum-rank-metric code $\CECode$, we have
    \begin{equation}\label{eq:dualSuppEqualErrorSupp}
    \SumRankSupp^\perp{(\Hce)} = \SumRankSupp{(\E)}.
    \end{equation}
\end{theorem}
\begin{IEEEproof}
First, partition $\Hce$ into blocks according to the length partition $\n$, i.e.,
\begin{equation}
     \Hce = \left[\Hce^{(1)}\mid\dots\mid\Hce^{(\shots)}\right]
\end{equation}
with $\Hce^{(i)}\in\Fqm^{(n-k-\errWeight)\times n_i}$ for all $i\inshots$.
We want to show that $\SumRankSupp^\perp{(\Hce)} = \SumRankSupp{(\E)}$. By the definition of the support for the sum-rank metric, this means that we need to show that
\begin{equation}
    \RankSupp^\perp{(\Hce^{(i)})} = \RankSupp{(\E^{(i)})} = \Rowspace{\B^{(i)}} \quad\forall i\inshots.
\end{equation}
Define $\mu_i \defeq \rkq{(\Hce^{(i)})}$ for all $i\inshots$. Then, $\Hce^{(i)}$ can be decomposed as
\begin{equation}
    \Hce^{(i)} = \C_{\CECode}^{(i)}\D_{\CECode}^{(i)}
\end{equation}
with $\C_{\CECode}^{(i)}\in\Fqm^{(n-k-\errWeight)\times \mu_i}$, $\D_{\CECode}^{(i)}\in\Fq^{\mu_i \times n_i}$, and $\rkq(\C_{\CECode}^{(i)}) = \rkq(\D_{\CECode}^{(i)}) = \mu_i$.

Recall from the definition of the sum-rank support~\eqref{eq:sumranksupport} and its dual support~\eqref{eq:dualSuppEqualErrorSupp} that we have
\begin{equation}
    \SumRankSupp^{\perp}{(\Hce)} = \Rowspace{\D_{\CECode}^{(1)}}^{\perp} \times \dots \times \Rowspace{\D_{\CECode}^{(\shots)}}^{\perp}
\end{equation}
and
\begin{equation}
    {\SumRankSupp{(\E)}} = \Rowspace{\B^{(1)}} \times \dots \times \Rowspace{\B^{(\shots)}},
\end{equation}
respectively.  
The goal is to show that $\Rowspace{\D_{\CECode}^{(i)}}^\perp = \Rowspace{\B^{(i)}}$ for all $i\inshots$, which is equivalent to proving $\Rowspace{\D_{\CECode}^{(i)}} = \Rowspace{\B^{(i)}}^\perp$. This will be achieved in two steps:
\begin{enumerate}
    \item Show that $\Rowspace{\D_{\CECode}^{(i)}} \subseteq \Rowspace{\B^{(i)}}^\perp$ for all $i\inshots$.
    \item Demonstrate that $\mu_i < \dim{(\Rowspace{\B^{(i)}}^\perp)} = n_i - \errWeight_i$ is not possible for any $i\inshots$, implying $\mu_i = n_i - \errWeight_i$ and hence $\Rowspace{\D_{\CECode}^{(i)}} = \Rowspace{\B^{(i)}}^\perp$ for all $i\inshots$.
\end{enumerate}
\medskip
\textbf{Step 1:}  Proving $\Rowspace{\D_{\CECode}^{(i)}} \subseteq \Rowspace{\B^{(i)}}^\perp$ for all $i\inshots$. \\
To prove $\Rowspace{\D_{\CECode}^{(i)}} \subseteq \Rowspace{\B^{(i)}}^\perp$, we instead show that $\Rowspace{\B^{(i)}} \subseteq \Rowspace{\D_{\CECode}^{(i)}}^\perp$.
By definition, $\Hce$ is a parity-check matrix for $\CECode = \errorCode + \CCode$. Thus,
\begin{equation}
    \Hce \G_{\errorCode}^\top = \bm{0} \quad\Leftrightarrow\quad \Hce \B^\top \A_{\errorCode}^\top = \bm{0}
\end{equation}
where $\G_{\errorCode}$ is the generator matrix of the error code as defined in~\eqref{eq:gen_error_code}.
Since $\A_{\errorCode}\in\Fqm^{\errWeight \times \errWeight}$ is non-singular, we have that
\begin{equation}\label{eq:BkernelZero}
    \Hce \B^\top = \bm{0} \Leftrightarrow \Hce^{(i)} {\B^{(i)}}^\top = \bm{0} \quad\forall i\inshots.
\end{equation}
This implies that all rows of $\B^{(i)}$ are in the $\Fqm$-right kernel of $\Hce^{(i)}$, and since $\B^{(i)}$ is over $\Fq$, we have that $\Rowspace{\B^{(i)}} \subseteq \Rowspace{\D_{\CECode}^{(i)}}^\perp$. Consequently, $\Rowspace{\D_{\CECode}^{(i)}} \subseteq \Rowspace{\B^{(i)}}^\perp$.

\medskip
\textbf{Step 2:} Showing that $\mu_i < \dim{(\Rowspace{\B^{(i)}}^\perp)} = n_i - \errWeight_i$ is impossible for any $i\inshots$. \\
Since $\Rowspace{\D_{\CECode}^{(i)}} \subseteq \Rowspace{\B^{(i)}}^\perp$, $\mu_i > n_i - \errWeight_i$ is not possible for any $i\inshots$. Assume that $\mu_{i'} < n_{i'} - \errWeight_{i'}$ for at least one $i'\inshots$, i.e., let $\mu_{i'} = n_{i'} - \errWeight_{i'} - \delta \in \ZZ$ with $\delta > 0$. Without loss of generality, set $i' = \shots$.

Given that $\rkq{(\D_{\CECode}^{(\shots)})} = n_\shots - \errWeight_\shots - \delta$, there exists a full-rank matrix $\Q^{(\shots)} \in \Fq^{n_\shots \times n_\shots}$ that allows us to bring $\D_{\CECode}^{(\shots)}$ into column-echelon form. Hence,
\begin{equation}
    \D_{\CECode}^{(\shots)} \Q^{(\shots)} = \left[\begin{array}{c|c}
    \smash{\underbrace{\mathbf{0}}_{\mathclap{\in\Fq^{(n_\shots-\errWeight_\shots-\delta) \times (\errWeight_\shots+\delta)}}}} & \widetilde{\D}_{\CECode}^{(\shots)}
    \end{array}\right] \\[1em]
\end{equation}
where $\widetilde{\D}_{\CECode}^{(\shots)} \in \Fq^{(n_\shots-\errWeight_\shots-\delta) \times (n_\shots-\errWeight_\shots-\delta)}$ with $\rkq{(\widetilde{\D}_{\CECode}^{(\shots)})} = n_\shots - \errWeight_\shots - \delta$.

Further, let 
\begin{equation}
    \Q^{(\shots)} = [\Q_1^{(\shots)} \mid \Q_2^{(\shots)}]
\end{equation}
with $\Q_1^{(\shots)} \in \Fq^{n_\shots \times (\errWeight_\shots + \delta)}$ and $\Q_2^{(\shots)} \in \Fq^{n_\shots \times (n_\shots-\errWeight_\shots - \delta)}$. Since $\Q^{(\shots)}$ is full-rank, we have that $\Q_1^{(\shots)}$ is full-rank too, i.e., $\rkq{(\Q_1^{(\shots)})} = \errWeight_\shots + \delta$. Thus,
\begin{equation}\label{eq:allmustbezero}
      \D_{\CECode}^{(\shots)} \Q_1^{(\shots)} = \mathbf{0}.
\end{equation}
That means we can multiply~\eqref{eq:allmustbezero} from the right with some full-rank transformation matrix $\T\in\Fq^{(\errWeight_\shots + \delta) \times (\errWeight_\shots + \delta)}$ such that
\begin{equation}\label{eq:allmustbezerov2}
      \D_{\CECode}^{(\shots)} \underbrace{\left[ {\B^{(\shots)\top}} \mid {\widetilde{\B}^{(\shots)\top}} \right]}_{=\Q_1^{(\shots)}\T} = \mathbf{0}.
\end{equation}

Define the following block-diagonal matrix
\begin{equation}
    \Q = \begin{bmatrix}
        \B^{(1)} & \mathbf{0} & \cdots &\mathbf{0} \\
        \mathbf{0} & \B^{(2)} & \cdots  &\mathbf{0} \\
        \vdots & \vdots & \ddots  & \vdots\\
        \mathbf{0} & \mathbf{0} & \cdots & \B^{(\shots)} \\
        \mathbf{0} & \mathbf{0} & \cdots & \widetilde{\B}^{(\shots)} \\
    \end{bmatrix} \in \Fq^{(\errWeight+\delta) \times n}.
\end{equation}
Then we have that
\begin{equation}\label{eq:allmustbezerov3}
    \D_{\CECode} \Q^\top = \bm{0} 
\end{equation}
since $\D_{\CECode}^{(i)} {\B^{(i)}}^\top = \bm{0}$ for $i\inshotsarg{\shots-1}$ and by assumption~\eqref{eq:allmustbezerov2}, $\D_{\CECode}^{(\shots)} \left[ {\B^{(\shots)\top}} \mid {\widetilde{\B}^{(\shots)\top}} \right] = \mathbf{0}$.

Now, without loss of generality, let $\delta = 1$. By the decoding condition~\eqref{eq:rankBandb}, we have that
\begin{equation}
    \rkqm{\left(\H \Q^\top\right)} = \errWeight + 1
\end{equation}
must hold. Thus, there exists a vector $\g \in \RowspaceFqm{\H}$ such that 
\begin{equation}
    \g \Q^\top = \begin{bmatrix} 0 & \dots & 0 & g_{\errWeight+1} \end{bmatrix} \neq \begin{bmatrix} 0 & \dots & 0 \end{bmatrix} \in \Fqm^{\errWeight+1}. 
\end{equation}   
Since the first $\errWeight$ leftmost positions of $\g \Q^\top$ are zero, by~\eqref{eq:BkernelZero} and the fact that the matrix formed by the $\errWeight$ leftmost columns in $\Q^\top$ forms a basis of all $\Rowspace{\B^{(i)}}^\perp$ for all $i\inshots$, which are also bases for $\RowspaceFqm{\B^{(i)}}^\perp$ for all $i\inshots$, this implies that $\g \in \RowspaceFqm{\B}^\perp$.

Also recall that $\Hce$ fulfills the parity-check constraints for both codes simultaneously: the error code $\errorCode$ and the component code $\CCode$. That means that
\begin{align}
\CECode = \CCode + \errorCode \Leftrightarrow \CECode^\perp &= \CCode^\perp \cap \errorCode^\perp \\
\Leftrightarrow \RowspaceFqm{\Hce}&= \RowspaceFqm{\H} \cap \RowspaceFqm{\B}^\perp.
\end{align}
Since for this specific $\g$ we have that $\g \in \RowspaceFqm{\H}$ and also $\g \in \RowspaceFqm{\B}^\perp$, it follows that $\g \in \RowspaceFqm{\Hce}$. Expanding $\g$ over $\Fq$ also implies that there exists a vector $\g' \in  \Rowspace{\Hce} =  \Rowspace{\D_{\CECode}}$ such that
\begin{equation}
    \g' \Q^\top =  \begin{bmatrix} 0 & \dots & 0 & g_{\errWeight+1}' \end{bmatrix} \neq \begin{bmatrix} 0 & \dots & 0 \end{bmatrix}  \in \Fq^{\errWeight+1}. %
\end{equation}
But by~\eqref{eq:allmustbezerov3}, for all $\g' \in \Rowspace{\D_{\CECode}}$ we need to have that
\begin{equation}
    \g' \Q^\top = \begin{bmatrix} 0 & \dots & 0 & 0 \end{bmatrix}  \in \Fq^{\errWeight+1}.
\end{equation}
This constitutes a contradiction, and thus $\mu_\shots < n_\shots - \errWeight_\shots$ is not possible. This also holds for any other $i' \neq \shots$, and therefore $\mu_i < n_i - \errWeight_i$ is not possible for any $i\inshots$. %

When $\delta=1$, we obtain one additional zero column in $\g' \Q^\top$. Similarly, when $\delta=2$, we get two additional zero columns. Since a contradiction arises for $\delta=1$, it follows that the assumption cannot hold for any $\delta > 1$ as well. For $\delta = 0$, we do not get a contradiction, and thus $\mu_i = n_i - \errWeight_i$ for all $i\inshots$ is the only valid option. %

This proves that $\Rowspace{\D_{\CECode}^{(i)}} =\Rowspace{\B^{(i)}}^\perp$ for all $i\inshots$, and therefore $\Rowspace{\D_{\CECode}^{(i)}}^\perp = \Rowspace{\B^{(i)}}$ for all $i\inshots$, hence $\SumRankSupp^\perp{(\Hce)} = \SumRankSupp{(\E)}$. %
\end{IEEEproof}

\begin{remark}
    Due to the properties of the error code and the relationship $\Y = \C + \E$, the following row spaces over $\Fqm$ are the same
    \begin{equation}
    \RowspaceFqm{\left[ \begin{array}{c}
    \G \\ \hline
    \G_\errorCode
    \end{array} \right] } = \RowspaceFqm{\left[ \begin{array}{c}
    \G \\ \hline
    \E
    \end{array} \right] } = \RowspaceFqm{\left[ \begin{array}{c}
    \G \\ \hline
    \Y
    \end{array} \right] }.
    \end{equation}
    Thus, the rows of all three matrices are generating sets for the code $\CECode = \CCode + \errorCode$.
\end{remark}

Note, that a parity-check matrix $\Hce$ for $\CECode$ can be obtained by stacking $\G_\CECode$ with $\Y$ and then performing Gaussian elimination.
This fact leads to the following observation for very high interleaving orders.

\begin{remark}%
    The error-code perspective on the Metzner--Kapturowski-like algorithm allows for new insights for very high-order interleaving orders, i.e., for $\intOrder \geq k + t$.
    In particular, if the rows of the transmitted codeword $\C$ form a generating set for $\CCode$, i.e., if $\rkqm(\C) = k$ and the error matrix $\E$ fulfills the full-rank condition, we have that $\rkqm(\Y) = k+t$ and the rows of $\Y$ form a generating set for $\CECode = \CCode + \errorCode$.

    This allows us to compute a parity-check matrix $\Hce$ for $\CECode$ directly from the received matrix $\Y$ as a basis for the right $\Fqm$-kernel of $\Y$ and recover the support of the error as $\SumRankSupp^\perp{(\Hce)} = \SumRankSupp{(\E)}$ (see~\eqref{eq:dualSuppEqualErrorSupp} in Theorem~\ref{thm:errorsupportinFqHdual}).
    Remarkably, we can recover the support of the error $\E$ without knowing the codes $\CCode$ and $\CECode$.
    
    This observation could be relevant for cryptosystems which rely on (secret) very high-order interleaved codes, since the knowledge of the error support could reduce the security level significantly, see e.g.~\cite{horlemann2021information}.
\end{remark}

We now present a theorem that establishes a direct connection between the syndrome matrix $\S$ and the parity-check matrix $\Hce$ of the sum code $\CECode = \CCode + \errorCode$. This theorem provides a straightforward method to compute $\Hce$ from $\S$ as used in the existing Metzner--Kapturowski variants for the Hamming and the rank metric.

\begin{theorem}\label{thm:get_Hsub}
    Let $\mycode{IC}[\intOrder;\n,k,d]$ be an $\Fqm$-linear interleaved sum-rank-metric code with component code $\CCode$, which has parity-check matrix $\H\in\Fqm^{(n-k)\times n}$. Let $\E\in\Fqm^{\intOrder\times n}$ be an error matrix with $\rkqm(\E)=\errWeight$ and $\SumRankWeight{\n}{(\E)}=\errWeight \leq n-k-1$ and let $\errorCode$ be the error code spanned by the rows of $\E$. The received word is $\Y = \C + \E$, where $\C\in\mycode{IC}$. The syndrome matrix is $\S = \H\Y^\top=\H\E^\top$, where $\S\in\Fqm^{(n-k)\times \intOrder}$.
    
    Let $\P\in\Fqm^{(n-k)\times (n-k)}$ be a full-rank matrix such that $\P\S$ is in row-echelon form, i.e.,
    \begin{equation}
        \P\S = \begin{bmatrix} \S' \\ \bm{0} \end{bmatrix} \longrightarrow \P\H = \begin{bmatrix} \H' \\ \Hce \end{bmatrix}
    \end{equation}
    where $\S'\in\Fqm^{\errWeight\times \intOrder}$, $\H'\in\Fqm^{\errWeight\times n}$ then $\Hce\in\Fqm^{(n-k-\errWeight)\times n}$ is a parity-check matrix for the sum-rank-metric code $\CECode = \errorCode + \CCode$ as defined in~\eqref{eq:codeanderrorsumcode}.
\end{theorem}
\begin{IEEEproof}
    Since $\P$ is invertible, multiplying both sides of $\S = \H\E^\top$ by $\P$ yields
    \begin{equation}
        \P\S = \P\H\E^\top.
    \end{equation}
    As $\H$ has full row rank $\rkqm(\H) = n-k$ and $\rkqm(\E) = t$, we have
    \begin{equation}
        \rkqm(\S) = \rkqm(\H\E^\top) = \min\{\rkqm(\H), \rkqm(\E)\} = \min\{n-k, t\} = t.
    \end{equation}
    By the rank-nullity theorem, $\rkqm(\P\S) = \rkqm(\S) = t$, so $\P\S$ has $t$ non-zero rows. As $\P\S$ is in row-echelon form, we can write
    \begin{equation}
        \P\S = \begin{bmatrix} \S' \\ \bm{0} \end{bmatrix},
    \end{equation}
    where $\S'\in\Fqm^{t\times s}$ has full row rank.
    
    Partitioning $\P\H$ conformally with $\P\S$, we have
    \begin{equation}
        \P\H = \begin{bmatrix} \H' \\ \Hce \end{bmatrix},
    \end{equation}
    where $\H'\in\Fqm^{t\times n}$ and $\Hce\in\Fqm^{(n-k-t)\times n}$. Since $\P\H\E^\top = \P\S$, we have
    \begin{equation}
        \begin{bmatrix} \H' \\ \Hce \end{bmatrix} \E^\top = \begin{bmatrix} \S' \\ \bm{0} \end{bmatrix},
    \end{equation}
    which implies $\Hce\E^\top = \bm{0}$. As the rows of $\E$ span $\errorCode$, this means $\Hce$ satisfies the parity-check equations for $\errorCode$. By construction, $\Hce$ also satisfies the parity-check equations for $\CCode$, as it is a submatrix of $\P\H$. And since $\Hce$ has $n-k-\errWeight$ rows and is of full-rank, it is a parity-check matrix for the sum-rank-metric code $\CECode$ defined in~\eqref{eq:codeanderrorsumcode}, which contains both $\CCode$ and $\errorCode$.
\end{IEEEproof}

\subsection{A \MeKap-like Decoding Algorithm}

Using Theorem~\ref{thm:errorsupportinFqHdual} and Theorem~\ref{thm:get_Hsub}, we can formulate an efficient decoding algorithm for high-order interleaved sum-rank-metric codes.
The algorithm is given in Algorithm~\ref{alg:decode_high_order_int_sum_rank} and proceeds similar to the \MeKap(-like) decoding algorithms for Hamming- or rank-metric codes.
As soon as $\Hsub$ is computed from the syndrome matrix $\S$, the rank support  of each block can be recovered independently using the results from Theorem~\ref{thm:errorsupportinFqHdual}.
This corresponds to finding a basis in the form of a matrix $\B^{(i)}\in\Fq^{t_i \times n_i}$ such that $\ext(\Hsub^{(i)})(\B^{(i)})^\top = \0$ for all $i\inshots$, where $t_i$ is determined by the rank-nullity theorem as $t_i = n_i - \rk_q(\Hsub^{(i)})$ according to \eqref{eq:dualSuppEqualErrorSupp}.

\begin{algorithm}[ht!]
    \setstretch{1.35}
    \caption{Decoding High-Order Interleaved Sum-Rank-Metric Codes}\label{alg:decode_high_order_int_sum_rank}
    \SetKwInOut{Input}{Input}\SetKwInOut{Output}{Output}
    
    \Input{Parity-check matrix $\H$ of $\mycode{C}$, Received word $\Y = \C + \E$ with $\C \in \mycode{IC}[\intOrder;\n,k,d]$ and $\SumRankWeight{\n}(\E) = \rkqm(\E) = t$}
    \Output{Transmitted codeword $\C$}
    \BlankLine
    
    $\S \gets \H\Y^\top \in \Fqm^{(n-k) \times \intOrder}$ \label{step:syndromeMatrix} \\
    Compute $\P\in\Fqm^{(n-k) \times (n-k)}$ s.t. $\P\S=\REF(\S)$ \label{step:transfP} \\
    $\Hsub=\left[\Hsub^{(1)} \sep \Hsub^{(2)} \sep \dots \sep \Hsub^{(\shots)} \right] \gets (\P\H)_{[t+1:n-k],[1:n]}\in\Fqm^{(n-t-k) \times n}$ \label{step:Hsub} \\
    \For{$i\inshots$}{
        Compute $\B^{(i)}\in\Fq^{t_i \times n_i}$ s.t. $\ext(\Hsub^{(i)})(\B^{(i)})^\top = \0$, where $t_i = n_i - \rk_q(\Hsub^{(i)})$ \label{step:getBi}
    }
    $\B \gets \diag(\B^{(1)}, \B^{(2)}, \dots, \B^{(\shots)})\in\Fq^{t \times n}$ \label{step:blockdiaB} \\
    Compute $\A\in\Fqm^{\intOrder \times t}$ s.t. $(\H\B^\top)\A^\top=\S$ \label{step:col_er_dec} \\
    $\C \gets \Y - \A\B\in\Fqm^{\intOrder \times n}$ \label{step:comp_E} \\
    \Return{$\C$} \label{step:return}
\end{algorithm}

\begin{theorem}
    Let $\C$ be a codeword of an $\intOrder$-interleaved sum-rank-metric code $\mycode{IC}[\intOrder;\n,k,d]$ and let $\H$ be the parity-check matrix of the corresponding constituent code $\mycode{C}$.
    Furthermore, let $\E \in \Fqm^{\intOrder \times n}$ be an error matrix of sum-rank weight $\SumRankWeight{\n}(\E)=t$ that fulfills $t\leq\intOrder$ (\emph{high-order condition})
    and $\rkqm(\E)=t$ (\emph{full-rank condition}). Let $\B$ be a basis of the $\Fq$-row space of $\E$. If~\eqref{eq:rankBandb} holds,
    then $\C$ can be uniquely recovered from the received word $\Y = \C + \E$ using Algorithm~\ref{alg:decode_high_order_int_sum_rank} in
    a time complexity equivalent to
    \begin{equation}
        \oh{\max\{n^3, n^2\intOrder\}}
    \end{equation}
    operations in $\Fqm$.
\end{theorem}

\begin{IEEEproof}
    Lemma~\ref{lem:col_erasure_decoding} states that the error matrix $\E$ can be factored as $\E = \A\B$.
    The decoding procedure in Algorithm~\ref{alg:decode_high_order_int_sum_rank} starts by finding a basis $\B$ of the error support $\SumRankSupp(\E)$ and then uses erasure decoding with respect to Lemma~\ref{lem:col_erasure_decoding} to recover $\A$.
    The matrix $\B$ is computed by transforming $\S$ into row-echelon form using a transformation matrix $\P$ (see Line~\ref{step:transfP}).
    In Line~\ref{step:Hsub}, $\Hsub$ is obtained by choosing the last $n-k-t$ rows of $\P\H$. 
    According to Theorem~\ref{thm:get_Hsub}, the matrix $\Hsub$ serves as a parity-check matrix for both the error code $\errorCode$ associated with the error matrix $\E$ and the component code $\mycode{C}$.
    Then using Theorem~\ref{thm:errorsupportinFqHdual} for each block (see Line~\ref{step:getBi}) we find a matrix $\B^{(i)}$ whose rows form a basis for ${\Rowspace{\ext(\Hsub^{(i)})}}^\top$ and therefore a basis for $\RankSupp(\E^{(i)})$ for all $i\inshots$.
    The matrix $\B$ is the block-diagonal matrix formed by $\B^{(i)}$ (cf.~\eqref{eq:def_B} and see Line~\ref{step:blockdiaB}) for $i\inshots$.
    Finally, $\A$ can be computed from $\B$ and $\H$ using Lemma~\ref{lem:col_erasure_decoding} in Line~\ref{step:col_er_dec}.
    Hence, Algorithm~\ref{alg:decode_high_order_int_sum_rank} returns the transmitted codeword in Line~\ref{step:return}.
    The complexities of the lines in the algorithm are as follows:
    \begin{itemize}
        \item Line~\ref{step:syndromeMatrix}: The syndrome matrix $\S = \H \Y^\top$ can be computed in at most $\oh{n^2\intOrder}$ operations in $\Fqm$.
        \item Line~\ref{step:transfP}: The transformation of $[\S \sep \I]$ into row-echelon form requires \begin{equation*}\oh{(n-k)^2(\intOrder+n-k)}\subseteq\oh{\max\{n^3, n^2\intOrder\}}\end{equation*} operations in $\Fqm$.
        \item Line~\ref{step:Hsub}: The product $(\P\H)_{[t+1:n-k],[1:n]}$ can be computed requiring at most 
        \begin{equation}
          \oh{n(n-k-t)(n-k)}\subseteq\oh{n^3} 
        \end{equation}
        operations in $\Fqm$.  
        \item Line~\ref{step:getBi}: The transformation of $[\ext(\Hsub^{(i)})^\top \sep \I^\top]^\top$ into column-echelon form requires $\oh{n_i^2 ((n-k-t)m+n_i)}$ operations in $\Fq$ per block. Overall we get 
        $$\oh{\sum_{i=1}^{\shots} n_i^2 ((n-k-t)m+n_i)}\subseteq\oh{n^3 m}$$
        operations in $\Fq$ since we have that $\oh{\sum_{i=1}^{\shots} n_i^2}\subseteq\oh{n^2}$.
        \item Line~\ref{step:col_er_dec}: According to Lemma~\ref{lem:col_erasure_decoding}, this step can be done in $\oh{(n-k)^3 m^2}$ operations over $\Fq$.
        \item Line~\ref{step:comp_E}: The product $\A\B = \left[\A^{(1)}\B^{(1)}\sep\A^{(2)}\B^{(2)}\sep\ldots\sep\A^{(\shots)}\B^{(\shots)}\right]$ can be computed in $\oh{\sum_{i=1}^{\shots}\intOrder t_i n_i} \subseteq \oh{\intOrder n^2}$ and the difference of $\Y - \A\B$ can be computed in $\oh{\intOrder n}$ operations in $\Fqm$. %
    \end{itemize}

    The complexities for {Line~\ref{step:getBi}} and Line~\ref{step:col_er_dec} are given for operations in $\Fq$.
    The number of $\Fq$-operations of both steps together is in $\oh{n^3 m^2}$ and their execution complexity can be bounded by $\oh{n^3}$ operations in $\Fqm$ (see~\cite{Couveignes2009}).

    Thus, Algorithm~\ref{alg:decode_high_order_int_sum_rank} requires $\oh{\max\{n^3, n^2\intOrder\}}$ operations in $\Fqm$.
\end{IEEEproof}

Note that the complexity of Algorithm~\ref{alg:decode_high_order_int_sum_rank} is not affected by the decoding complexity of the underlying constituent code since a generic code with no structure is assumed.

\section{Further Results and Remarks}

\subsection{Probabilistic Decoding for Uniform Random Errors}

In practical settings, the full-rank condition may not always hold. Therefore, we consider the performance of the decoder when the error is drawn uniformly at random from the set of all error matrices of a given sum-rank weight $\errWeight$. We then derive an upper bound on the error probability, which, for fixed code parameters, decays exponentially with respect to the difference between the error weight $\errWeight$ and the interleaving order $\intOrder$.

Note that we still require the high-order condition, i.e., $\intOrder \geq \errWeight$. Otherwise, no error can possibly satisfy the full-rank condition since
\begin{equation}
    \rkqm(\E) \leq \sum_{i=1}^{\shots} \rkqm(\shot{\E}{i})
    \leq \sum_{i=1}^{\shots} \rkq(\shot{\E}{i}) = \sum_{i=1}^{\shots} \errWeight_i = \errWeight
\end{equation}
holds, and $\E$ has size $\intOrder \times n$ (with $\intOrder \leq n$).

For the sake of simplicity in the analysis, we focus on the case where the length partition $\n = [n_1, \dots, n_\shots]$ has constant block lengths, i.e., there exists a positive integer $\eta$ such that $n_i=\eta$ for all $i\inshots$.

We introduce the following sets, which are integral to the proofs of the forthcoming theorems in this section.

First define $\mu$ as the maximum possible $\Fq$-rank of each block of the error matrix, given by
\begin{equation}\label{eq:defmu}
    \mu \defeq \min\{sm, \eta\}.
\end{equation}

Next, we define the set of all possible rank profiles $\errWeightVec = [\errWeight_1, \dots, \errWeight_\shots]$ for any error matrix  with $\shots$ blocks and sum-rank weight $\errWeight$, where each component $\errWeight_i$ is bounded by $\mu$ as
\begin{equation}
    \wdecomp{\errWeight,\shots,\mu} \defeq \left\{ \errWeightVec \in \{0,\ldots,\mu\}^\shots : \sum_{i=1}^{\shots} \errWeight_i = \errWeight \right\}.
\end{equation}
This set will be used to enumerate all possible rank profiles.

For a given length partition $\n$, we define the set of all error matrices with sum-rank weight $\errWeight$ as follows
\begin{equation}
    \mathcal{E}_{\errWeight}^{(\n)} \defeq \left\{ \E=\left[\E^{(1)}\mid\dots\mid\E^{(\shots)}\right] \in  \Fqm^{\intOrder  \times n}  : \SumRankWeight{\n}(\E) = \sum_{i=1}^{\shots} \rkq(\E^{(i)}) = \errWeight \right\}.
\end{equation}
This set contains all possible error matrices with the specified sum-rank weight $\errWeight$ and length partition $\n$.
For a fixed rank profile we define
\begin{equation}
    \mathcal{E}_{\errWeightVec}^{(\n)} \defeq \left\{ \E=\left[\E^{(1)}\mid\dots\mid\E^{(\shots)}\right] \in  \Fqm^{\intOrder  \times n}  : \rkq(\E^{(i)}) = \errWeight_i \right\}.
\end{equation}
and from~\eqref{eq:error_decomp} we have that we can decompose the error into $\E = \A\B$ with $\A\in\Fqm^{\intOrder \times \errWeight}$ and $\B\in\Fq^{\errWeight \times n}$ with $\A$ and $\B$ both of full-rank.
Let us define the set of all possible matrices $\A$
\begin{equation}\label{eq:def_A_Set}
 \set{A}_{\errWeightVec} \defeq \left\{\A\in\Fqm^{\intOrder\times \errWeight} : \SumRankWeight{\errWeightVec}(\A)=\errWeight\right\}   
\end{equation}
and all possible matrices $\B$ as
\begin{equation}\label{eq:def_B_Set}
    \set{B}_{\errWeightVec} \defeq \left\{\diag(\B^{(1)},\ldots,\B^{(\shots)})\in\Fq^{\errWeight\times n} : \rkq{(\B^{(i)})}=\errWeight_i \text{ and } \B^{(i)}\in\Fq^{\errWeight_i \times \eta} \;\forall i\inshots\right\}.
\end{equation}

When drawing $\E$ uniformly at random from $\mathcal{E}_{\errWeight}^{(\n)}$ the marginal distribution for the corresponding rank profile $\errWeightVec\in\wdecomp{\errWeight,\shots,\mu}$ is given by
\begin{equation}
    \Pr[\errWeightVec] = \frac{1}{|\set{E}_\errWeight^{(\n)}|} \prod_{i=1}^{\shots}\NM_q(\intOrder m,\eta,\errWeight_i)
\end{equation}
where $\NM_q(\intOrder m,\eta,\errWeight_i)$ denotes the number of matrices over $\Fq$ of size $\intOrder m \times \eta$ of rank $\errWeight_i$ which can be computed as~(see~\cite{puchingerGenericDecodingSumRank2022})
\begin{equation}
    \NM_q(\intOrder m,\eta,\errWeight_i) = \prod_{j=0}^{\errWeight_i - 1} \frac{(q^{\intOrder m}-q^j)(q^\eta - q^j)}{q^{\errWeight_i} - q^j}.
\end{equation}

\begin{lemma}\label{lem:probLemma1}
    For a given rank profile $\t = [t_1, t_2, \ldots, t_\shots ]$ of the error $\E$, 
    the probability that $\E$ has $\Fqm$-rank equal to $t$, given $\t$ is then
    \begin{eqnarray}
        \Pr[\rkqm(\E) = t  \,|\, \t] =& \Pr[\rkqm(\A) = t \,|\, \t] \\
                                     =& \frac{\prod_{j=0}^{t-1}(q^{sm}-q^{jm})}{\prod_{i=1}^{\shots}\prod_{j=0}^{t_i-1}(q^{sm}-q^{j})} 
    \end{eqnarray}
    where $\A$ is a matrix drawn uniformly at random from the set defined in~\eqref{eq:def_A_Set}.
\end{lemma}
\begin{IEEEproof}
Every error matrix $\E$ can be decomposed as in~\eqref{eq:error_decomp}, i.e., $\E=\A\B$. Since $\A$ is the only part influencing the $\Fqm$-rank of $\E$ and is unique if an arbitrary block-diagonal matrix $\B$ with $\Rowspace{\B} = \Rowspace{\E}$ is fixed (see, e.g., \cite[Theorem 1]{matsaglia_equalities_1974}), we obtain
\begin{equation}
\Pr[\rkqm(\E)=t \,|\, \t] = \Pr[\rkqm(\A)=t \,|\, \t].
\end{equation}

Recall that $\set{B}_{\t}$ is defined in~\eqref{eq:def_B_Set} as the set of all block-diagonal matrices $\B$ with $\Rowspace{\B} = \Rowspace{\E}$ and rank profile $\t$. By the law of total probability, we then have
\begin{align}
    \Pr[\rkqm(\E)=t \,|\, \t] &= \sum_{\B \in \set{B}_{\t}} \Pr[\rkqm(\A) = t \,|\, \t,\B] \cdot \Pr[\B \,|\, \t] \\
    &= \sum_{\B \in \set{B}_{\t}} \Pr[\rkqm(\A) = t \,|\, \t] \cdot \Pr[\B \,|\, \t] \\
    &= \Pr[\rkqm(\A) = t \,|\, \t],
\end{align}
where we used the fact that $\Pr[\B \,|\, \t] = \frac{1}{|\set{B}_{\t}|}$ since $\B$ is uniformly distributed over $\set{B}_{\t}$.
The probability $\Pr[\rkqm(\A) = t\,|\, \t]$ can be computed as
\begin{equation}
    \Pr[\rkqm(\A)=t | \t] = \frac{|\{\A'\in\Fqm^{s\times t} : \SumRankWeight{\t}{(\A')}=\rkqm(\A')=t\}|}{|\{\A'\in\Fqm^{s\times t} : \SumRankWeight{\t}(\A')=t\}|}.
\end{equation}

Consider any matrix $\A\in\{\A'\in\Fqm^{\intOrder\times \errWeight} : \rkqm(\A')=\errWeight\}$. Since $\A$ is full-rank over $\Fqm$ and $\intOrder \geq \errWeight$, we can make several observations about the ranks of its blocks $\A^{(i)}$. First, the $\Fqm$-rank of each block $\A^{(i)}$ is equal to its corresponding rank profile component, i.e., $\rkqm(\A^{(i)}) = \errWeight_i$. Moreover, the $\Fq$-rank of each block $\A^{(i)}$ is lower bounded by its $\Fqm$-rank, meaning that $\errWeight_i \leq \rkq(\A^{(i)})$. At the same time, the $\Fq$-rank of each block $\A^{(i)}$ is upper bounded by $\min(\errWeight_i, \intOrder)$, because the rank of a matrix cannot exceed its number of rows or columns. In this case, each block $\A^{(i)}$ has dimensions $\intOrder \times \errWeight_i$, so its $\Fq$-rank is at most $\min\{\errWeight_i, \intOrder\}$. However, since $\errWeight = \sum_{i=1}^{\shots} \errWeight_i \leq \intOrder$, we have $\errWeight_i \leq \intOrder$ for all $i \inshots$, which implies that $\min\{\errWeight_i, \intOrder\} = \errWeight_i$. By combining the lower and upper bounds, we conclude that $\rkq{(\A^{(i)})}=\errWeight_i$ for all $i \inshots$.
This implies that for any $\A\in\{\A'\in\Fqm^{\intOrder\times \errWeight} : \rkqm(\A')=\errWeight\}$ we have that $\SumRankWeight{\errWeightVec}(\A)=\errWeight$ and therefore, we have the following equality
\begin{equation}
    \{\A'\in\Fqm^{\intOrder\times \errWeight} : \SumRankWeight{\errWeightVec}{(\A')}=\rkqm(\A')=\errWeight\}=\{\A'\in\Fqm^{\intOrder\times \errWeight} : \rkqm(\A')=\errWeight\}
\end{equation}
and hence
\begin{align}
    \Pr[\rkqm(\A)=t | \t] = \frac{|\{\A'\in\Fqm^{s\times t} : \rkqm(\A')=t\}|}{|\{\A'\in\Fqm^{s\times t} : \SumRankWeight{\t}(\A')=t\}|} = \frac{\prod_{j=0}^{t-1}(q^{sm}-q^{jm})}{\prod_{i=1}^{\shots}\prod_{j=0}^{t_i-1}(q^{sm}-q^{j})}
\end{align}
where $\prod_{j=0}^{t-1}(q^{sm}-q^{jm})$ is the number of all full-rank matrices of size $\intOrder\times\errWeight$ over $\Fqm$ and $\prod_{i=1}^{\shots}\prod_{j=0}^{t_i-1}(q^{sm}-q^{j})$ is the number of all matrices in $\Fqm^{s\times\errWeight}$ with sum-rank weight $\errWeight$ with corresponding length partition $\errWeightVec$~(see~\cite{puchingerGenericDecodingSumRank2022}).
\end{IEEEproof}

\begin{lemma}\label{lem:successProb}
    Let $\E$ be an error matrix drawn uniformly at random from the set $\set{E}_\errWeight^{(\n)}$. Then, the probability that $\rkqm(\E)=\errWeight$ is given by
    \begin{equation}\label{eq:succ_probability}
        \Pr[\rkqm(\E)=\errWeight] = \frac{\prod_{j=0}^{\errWeight-1}(q^{\intOrder m}-q^{jm})}{|\set{E}_\errWeight^{(\n)}|} \cdot\sum_{\errWeightVec \in \wdecomp{\errWeight,\shots,\mu}} \prod_{i=1}^{\shots} \quadbinom{\eta}{\errWeight_i}.
    \end{equation}
\end{lemma}
\begin{IEEEproof}
Recall the sets $\set{A}_{\errWeightVec}$ and $\set{B}_{\errWeightVec}$ defined in \eqref{eq:def_A_Set} and \eqref{eq:def_B_Set}, respectively.

According to Lemma~\ref{lem:probLemma1}, for a fixed rank profile $\errWeightVec$, we can draw $\A\in\set{A}_{\errWeightVec}$ and $\B\in\set{B}_{\errWeightVec}$ independently and uniformly from their corresponding domains and obtain $\E=\A\B$ with $\SumRankWeight{\n}(\E)=\errWeight$ such that $\E$ is uniformly drawn at random from $\set{E}_{\errWeightVec}$. 

This means the probability $\Pr[\rkqm(\E)=\errWeight]$ is
\begin{align}
    \Pr[\rkqm(\E)=\errWeight] &= \sum_{\errWeightVec \in \wdecomp{\errWeight,\shots,\mu}} \Pr[\errWeightVec]\cdot\Pr[\rkqm(\A)=\errWeight \;|\; \errWeightVec] \\
    &= \sum_{\errWeightVec \in \wdecomp{\errWeight,\shots,\mu}} \frac{\prod_{i=1}^{\shots}\NM_q(\intOrder m,\eta,\errWeight_i)}{|\set{E}_\errWeight^{(\n)}|} \cdot \frac{\prod_{j=0}^{\errWeight-1}(q^{\intOrder m}-q^{jm})}{\prod_{i=1}^{\shots}\prod_{j=0}^{\errWeight_i-1}(q^{\intOrder m}-q^{j})} \\
    &= \frac{\prod_{j=0}^{\errWeight-1}(q^{\intOrder m}-q^{jm})}{|\set{E}_\errWeight^{(\n)}|} \cdot\sum_{\errWeightVec \in \wdecomp{\errWeight,\shots,\mu}} \frac{\prod_{i=1}^{\shots}\NM_q(\intOrder m,\eta,\errWeight_i)}{\prod_{i=1}^{\shots}\prod_{j=0}^{\errWeight_i-1}(q^{\intOrder m}-q^{j})}.
\end{align}
Here, we first apply the law of total probability to express $\Pr[\rkqm(\E)=\errWeight]$ as a sum over all possible rank profiles $\errWeightVec \in \wdecomp{\errWeight,\shots,\mu}$. Then, we use the fact that $\A$ and $\B$ are drawn independently and uniformly from their respective domains to compute the conditional probability $\Pr[\rkqm(\A)=\errWeight \;|\; \errWeightVec]$.

Next, we simplify the expression using the definition of the Gaussian binomial coefficient:
\begin{align}
    \Pr[\rkqm(\E)=\errWeight] &= \frac{\prod_{j=0}^{\errWeight-1}(q^{\intOrder m}-q^{jm})}{|\set{E}_\errWeight^{(\n)}|} \cdot\sum_{\errWeightVec \in \wdecomp{\errWeight,\shots,\mu}} \frac{\prod_{i=1}^{\shots} \prod_{j=0}^{\errWeight_i-1}\frac{(q^{\intOrder m}-q^{j})(q^{\eta}-q^{j})}{(q^{\errWeight_i}-q^{j})}}{\prod_{i=1}^{\shots}\prod_{j=0}^{\errWeight_i-1}(q^{\intOrder m}-q^{j})}  \\
    &= \frac{\prod_{j=0}^{\errWeight-1}(q^{\intOrder m}-q^{jm})}{|\set{E}_\errWeight^{(\n)}|} \cdot\sum_{\errWeightVec \in \wdecomp{\errWeight,\shots,\mu}} {\prod_{i=1}^{\shots} \prod_{j=0}^{\errWeight_i-1}\frac{(q^{\eta}-q^{j})}{(q^{\errWeight_i}-q^{j})}}  \\
    &= \frac{\prod_{j=0}^{\errWeight-1}(q^{\intOrder m}-q^{jm})}{|\set{E}_\errWeight^{(\n)}|} \cdot\sum_{\errWeightVec \in \wdecomp{\errWeight,\shots,\mu}} \prod_{i=1}^{\shots} \quadbinom{\eta}{\errWeight_i}.
\end{align}
In the first step, we rewrite the numerator using the definition of $\NM_q(\intOrder m,\eta,\errWeight_i)$.  Then, we cancel out the common terms in the numerator and denominator, leaving only the Gaussian binomial coefficients in the final expression, which completes the proof.
\end{IEEEproof}

At first glance, the expression in~\eqref{eq:succ_probability} does not appear to be computationally efficient. However, in~\cite{puchinger2020generic}, it was shown that the term $|\set{E}_\errWeight^{(\n)}|$ can be efficiently computed using a dynamic programming approach. Inspired by this, we propose a similar procedure to compute the right-hand side of~\eqref{eq:succ_probability}. To this end, let us define
\begin{equation}
    \Phi_{q,\eta}(\errWeight, \shots) \defeq \sum_{\t \in \set{T}_{\errWeight,\shots,\mu}} \prod_{i=1}^{\shots} \quadbinom{\eta}{\errWeight_i}
\end{equation}
where $\Phi_{q,\eta}(\errWeight, \shots) $ represents the sum over all possible rank profiles $\t$ for a given sum-rank weight $\errWeight$. For each rank profile, the q-binomial coefficient $\quadbinom{\eta}{\errWeight_i}$ counts the number of subspaces of dimension $\errWeight_i$ in an $\eta$-dimensional space over $\Fq$.
This expression can be computed recursively as
\begin{equation}\label{eq:recursive_phi}
    \Phi_{q,\eta}(\errWeight, \shots) = \begin{cases}
        \displaystyle\quadbinom{\eta}{\errWeight} & \text{if } \shots = 1 \\
        \displaystyle\sum_{\errWeight'=0}^{\min\{\eta, \errWeight\}} \quadbinom{\eta}{\errWeight'} \cdot \Phi_{q,\eta}(\errWeight-\errWeight', \shots-1)& \text{else}
    \end{cases}.
\end{equation}
The recursive relation can be understood as follows: For the base case, when $\shots = 1$, there is only one block, and the number of subspaces of dimension $\errWeight$ in an $\eta$-dimensional space over $\Fq$ is given by the q-binomial coefficient $\quadbinom{\eta}{\errWeight}$. For $\shots > 1$, we consider all possible dimensions $\errWeight'$ for the first block, ranging from $0$ to $\min\{\eta, \errWeight\}$. For each choice of $\errWeight'$, we multiply the number of subspaces of dimension $\errWeight'$ in the first block, given by $\quadbinom{\eta}{\errWeight'}$, with the number of ways to distribute the remaining sum-rank weight $\errWeight-\errWeight'$ among the remaining $\shots-1$ blocks, recursively computed by $\Phi_{q,\eta}(\errWeight-\errWeight', \shots-1)$.

\begin{algorithm}[ht!]
    \setstretch{1.35}
    \caption{Compute $\Phi_{q,\eta}(\errWeight, \shots)$}\label{alg:compute_phi}
    \SetKwInOut{Input}{Input}
    \SetKwInOut{Output}{Output}
    \SetKwInOut{Initialize}{Initialize}
    
    \Input{Parameters: $q, \eta, \errWeight$ and $\shots$}
    \Output{$\Phi_{q,\eta}(\errWeight, \shots)$}
    \Initialize{$N(\errWeight', \shots') = 0 \quad \forall \errWeight'\inshotsarg{\errWeight}$ and $\shots'\inshots$}
    \For{$\errWeight'\inshotsarg{\errWeight}$}{
        $N(\errWeight', 1) \gets \quadbinom{\eta}{\errWeight'}$
    }
    \For{$\shots'\in\{2,\ldots,\shots\}$}{
        \For{$\errWeight'\inshotsarg{\errWeight}$}{
            $N(\errWeight', \shots') \gets \sum_{\errWeight'' = 0}^{\min\{\eta, \errWeight'\}} N(\errWeight'-\errWeight'', \shots'-1)  \cdot \quadbinom{\eta}{\errWeight''}$
        }
    }
\Return{$N(\errWeight,\shots)$}
\end{algorithm}

\begin{theorem}\label{thm:compute_phi_complexity}
    Algorithm~\ref{alg:compute_phi} is correct and requires $\shots \cdot \errWeight^2$ integer multiplications.
\end{theorem}
\begin{IEEEproof}
The correctness of Algorithm~\ref{alg:compute_phi} follows from the recursive relationship established in~\eqref{eq:recursive_phi} with the base cases $\Phi_{q,\eta}(\errWeight, 1) = \quadbinom{\eta}{\errWeight}$. 

Regarding the complexity, the algorithm performs $\shots \cdot \errWeight^2$ integer multiplications. This is because, for each $\shots' \in \{2, \ldots, \shots\}$ and each $\errWeight' \inshotsarg{\errWeight}$, the inner loop runs over $\min\{\eta, \errWeight'\}$ values, leading to at most $\errWeight$ iterations per combination of $\shots'$ and $\errWeight'$. Thus, the total number of iterations is $\shots \cdot \errWeight^2$.
\end{IEEEproof}

\begin{corollary}
    The success probability in~\eqref{eq:succ_probability} can be computed with polynomially-bounded complexity.
\end{corollary}
\begin{IEEEproof}
    The success probability in~\eqref{eq:succ_probability} is given by
    \begin{equation*}
        \frac{\prod_{j=0}^{\errWeight-1}(q^{\intOrder m}-q^{jm})}{|\set{E}_\errWeight^{(\n)}|} \cdot \sum_{\errWeightVec \in \wdecomp{\errWeight,\shots,\mu}} \prod_{i=1}^{\shots} \quadbinom{\eta}{\errWeight_i}.
    \end{equation*}
    We analyze the complexity of computing each term in this expression:
    \begin{itemize}
        \item $|\set{E}_\errWeight^{(\n)}|$ can be computed with polynomially bounded complexity, as shown in~\cite{puchingerGenericDecodingSumRank2022}.
        \item $\sum_{\errWeightVec \in \wdecomp{\errWeight,\shots,\mu}} \prod_{i=1}^{\shots} \quadbinom{\eta}{\errWeight_i}$ can be computed with polynomially bounded complexity according to Theorem~\ref{thm:compute_phi_complexity}.
        \item The computation of $\prod_{j=0}^{\errWeight-1}(q^{\intOrder m}-q^{jm})$ is also polynomially bounded. The terms $q^{\intOrder m}$ and $q^{jm}$ can be computed using repeated squaring, and their differences and products involve polynomially-bounded integer operations.
    \end{itemize}
The overall complexity is dominated by the complexity of computing $|\set{E}_\errWeight^{(\n)}|$ and the term $\sum_{\errWeightVec \in \wdecomp{\errWeight,\shots,\mu}} \prod_{i=1}^{\shots} \quadbinom{\eta}{\errWeight_i}$, both of which are polynomially bounded. Thus, the success probability can be computed with polynomially bounded complexity.
\end{IEEEproof}

\begin{theorem}\label{thm:failure_bound}
    Let $\mycode{IC}[\intOrder;\n,k,\dmin]$ be an $\Fqm$-linear homogeneous $\intOrder$-interleaved sum-rank-metric code with component code $\mycode{C}$ of minimum sum-rank distance $\dmin$, and let $\errWeight \leq \min\{\intOrder, \dmin-2\}$.
    Furthermore, let
    \begin{equation}
        \Y = \C + \E
    \end{equation}
    where $\C$ is a codeword of the interleaved code $\mycode{IC}[\intOrder;\n,k,\dmin]$ and $\E \in \Fqm^{\intOrder \times n}$ is an error matrix uniformly drawn at random from $\mathcal{E}_{\errWeight}^{(\n)}$. Then the probability that Algorithm~\ref{alg:decode_high_order_int_sum_rank} cannot decode, which is the probability that $\rkqm(\E) \neq \errWeight$, is bounded from above as
    \begin{equation}\label{eq:failure_bound}
        \Pr[\rkqm(\E) \neq \errWeight] \leq \errWeight q^{-m(\intOrder-\errWeight+1)}.
    \end{equation}
\end{theorem}
\begin{IEEEproof}
    From Lemma~\ref{lem:successProb}, we have
    \begin{align}
        \Pr[\rkqm(\E)=\errWeight] &= \frac{\prod_{j=0}^{\errWeight-1}(q^{\intOrder m}-q^{jm})}{|\set{E}_{\errWeight}^{(\n)}|} \cdot \sum_{\errWeightVec \in \wdecomp{\errWeight,\shots,\mu}} \prod_{i=1}^{\shots} \quadbinom{\eta}{t_i}.
    \end{align}
    
    Next, we consider the following inequality to bound the denominator \(|\set{E}_{\errWeight}^{(\n)}|\)
    \begin{align}
        |\set{E}_{\errWeight}^{(\n)}| &= \sum_{\errWeightVec \in \wdecomp{\errWeight,\shots,\mu}} \prod_{i=1}^{\shots} \quadbinom{\eta}{t_i} \prod_{j=0}^{t_i - 1}(q^{\intOrder m}-q^j) \\
        &\leq \sum_{\errWeightVec \in \wdecomp{\errWeight,\shots,\mu}} \prod_{i=1}^{\shots} \quadbinom{\eta}{t_i} \prod_{j=0}^{t_i - 1}q^{\intOrder m} \\
        &= \sum_{\errWeightVec \in \wdecomp{\errWeight,\shots,\mu}} \left( \prod_{i=1}^{\shots} \quadbinom{\eta}{t_i} \right) q^{\intOrder m \errWeight}.
    \end{align}

    Using this inequality, we can further bound \(\Pr[\rkqm(\E)=\errWeight]\) as follows
    \begin{align}
        \Pr[\rkqm(\E)=\errWeight] &\geq \frac{\prod_{j=0}^{\errWeight-1}(q^{\intOrder m}-q^{jm})}{q^{\intOrder m \errWeight}} \\ %
        &= \prod_{j=0}^{\errWeight-1}(1-q^{m(j-\intOrder)}) \geq 1 - \errWeight q^{m(\errWeight-\intOrder-1)}.
    \end{align}

    At this point, we have the same equation as in the rank-metric case. The last step follows from \cite[Theorem 10]{renner2021decoding}. 
    
    Finally, the claim of the theorem follows from the fact that 
    \begin{equation}
     \Pr[\rkqm(\E) \neq \errWeight] = 1 - \Pr[\rkqm(\E) = \errWeight].
    \end{equation}
\end{IEEEproof}

\begin{figure}[ht]
  \centering
  
  \begin{subfigure}{0.48\textwidth}
    \centering
    \begin{tikzpicture}
      \begin{axis}[
          width=\linewidth, %
          height=8cm,
          grid=major, 
          grid style={dotted,gray!80},
          xlabel={$\intOrder - \errWeight$}, %
          ylabel={Log Failure Probability (Base 10)},
          xmax=3,
          xmin=0,
          ymax=0,
          ymin=-2.5,
          xtick=data,
          tick label style={font=\scriptsize},
          legend style={font=\scriptsize}, 
          label style={inner sep=0, font=\small},
        ]
        
        \addplot[line1, mark=*, mark options={solid}] 
        table[x=st,y=logprobability_base10,col sep=comma] {./data/mk_succ_n10_q2_m2_t4_L1.txt}; 
        \addplot[line2, mark=square*, mark options={solid, fill=white}] 
        table[x=st,y=logprobability_base10,col sep=comma] {./data/mk_succ_n10_q2_m2_t4_L5.txt}; 
        \addplot[line3, mark=triangle*, mark options={solid}] 
        table[x=st,y=logprobability_base10,col sep=comma] {./data/mk_succ_n10_q2_m2_t4_L10.txt}; 
        \addplot[line4] 
        table[x=st,y=bound_log_base10,col sep=comma] {./data/mk_bound_n10_q2_m2_t4.txt}; 

        \legend{$\shots=1$, $\shots=5$, $\shots=10$, Bound~\eqref{eq:failure_bound}}    
      \end{axis}
    \end{tikzpicture}
    \caption{$q=2$, $m=2$, $n=10$ and $t=4$}
    \label{fig:probability_plot_1}
  \end{subfigure}
  \hfill
  \begin{subfigure}{0.48\textwidth}
    \centering
    \begin{tikzpicture}
      \begin{axis}[
          width=\linewidth, %
          height=8cm,
          grid=major, 
          grid style={dotted,gray!80},
          xlabel={$\intOrder - \errWeight$}, %
          ylabel={Log Failure Probability (Base 10)},
          xmax=10,
          xmin=0,
          ymax=0,
          ymin=-30,
          xtick=data,
          tick label style={font=\scriptsize},
          legend style={font=\scriptsize}, 
          label style={inner sep=0, font=\small},
        ]
        
        \addplot[line1, mark=*, mark options={solid}] 
        table[x=st,y=logprobability_base10,col sep=comma] {./data/mk_succ_n30_q2_m10_t11_L1.txt}; 
        \addplot[line2, mark=square*, mark options={solid, fill=white}] 
        table[x=st,y=logprobability_base10,col sep=comma] {./data/mk_succ_n30_q2_m10_t11_L6.txt}; 
        \addplot[line3, mark=triangle*, mark options={solid}] 
        table[x=st,y=logprobability_base10,col sep=comma] {./data/mk_succ_n30_q2_m10_t11_L30.txt}; 
        \addplot[line4] 
        table[x=st,y=bound_log_base10,col sep=comma] {./data/mk_bound_n30_q2_m10_t11.txt}; 

        \legend{$\shots=1$, $\shots=6$, $\shots=30$, Bound~\eqref{eq:failure_bound}}    
      \end{axis}
    \end{tikzpicture}
    \caption{$q=2$, $m=10$, $n=30$ and $t=11$}
    \label{fig:probability_plot_2}
  \end{subfigure}
  
  \caption{Logarithmic failure probability vs. $\intOrder-\errWeight$ for different values of $\shots$ with $q$, $m$, $n$ and $t$.}
  \label{fig:probability_plots}
\end{figure}
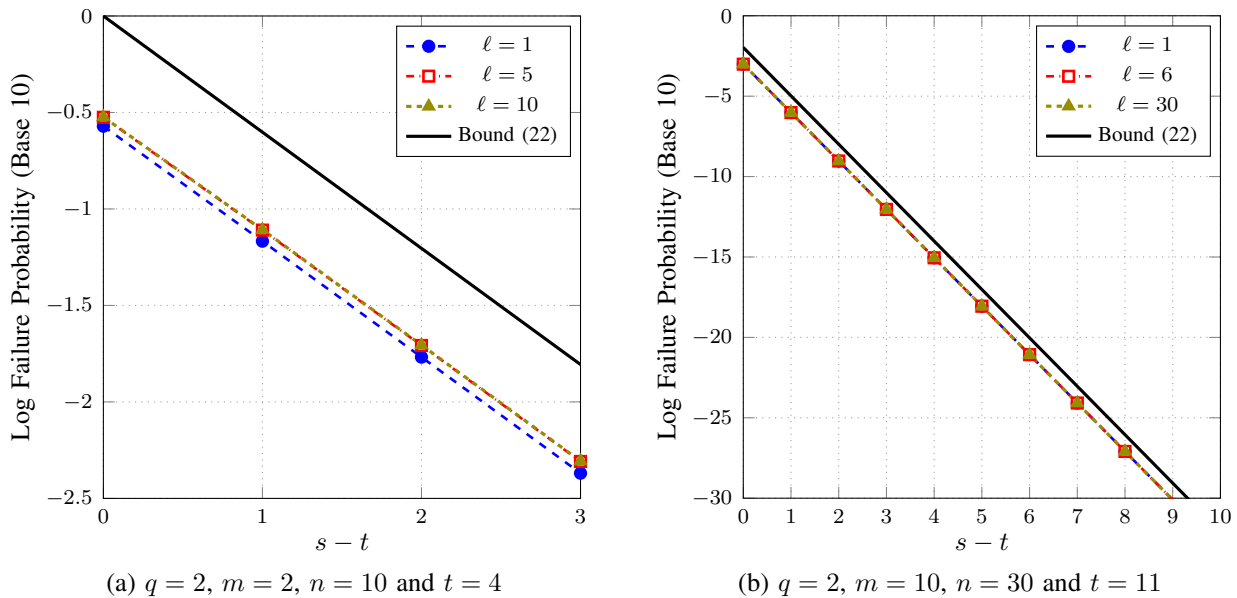

In Figure~\ref{fig:probability_plots}, we show the actual value of the failure probability, using Algorithm~\ref{alg:compute_phi} to evaluate~\eqref{eq:succ_probability} and compare with the derived upper bound from~\eqref{eq:failure_bound}. The failure probability is presented in logarithmic scale (base 10) versus the difference between the interleaving order $\intOrder$ and the sum-rank error weight $\errWeight$ for two different parameter sets.

Figure~\ref{fig:probability_plot_1} illustrates the failure probability for very small code parameters, with $q=2$, $m=2$, $n=10$, and $t=4$. On the other hand, Figure~\ref{fig:probability_plot_2} shows the failure probability for larger, but still relatively small, code parameters, with $q=2$, $m=10$, $n=30$, and $t=11$. 

From these plots, we can observe several key points:
\begin{enumerate}
    \item As the code parameters increase, the difference in failure probability between the rank metric ($\shots=1$), sum-rank metric ($1 < \shots < n$), and Hamming metric ($\shots=n$) becomes negligibly small. This suggests that for sufficiently large code parameters, the choice of metric has a diminishing impact on the failure probability.

    \item The failure probability declines exponentially fast as $\intOrder - \errWeight$ increases, which is expected based on the expression of the upper bound in~\eqref{eq:failure_bound}.

    \item The gap between the upper bound and the actual failure probability narrows as the code parameters increase. In Figure~\ref{fig:probability_plot_2}, with larger code parameters, the bound and the actual values are more closely aligned compared to Figure~\ref{fig:probability_plot_1}. This suggests that the derived upper bound becomes tighter and more accurate for larger code parameters.
\end{enumerate}

\subsection{Decoding Radius}

For the decoder presented in Algorithm~\ref{alg:decode_high_order_int_sum_rank} to succeed and uniquely recover the error, the following conditions must be satisfied:
\begin{enumerate}
    \item The error matrix $\E$ must satisfy the high-order and full-rank conditions, i.e., $\intOrder \geq \errWeight$ and $\rkqm(\E)=\errWeight$. Note that the full-rank condition already implies a high interleaving order, since for $\E$ to have rank $\errWeight$, the interleaving order $\intOrder$ must be at least $\errWeight$.
    \item The parity-check matrix $\H$ must satisfy the condition in~\eqref{eq:rankBandb}, which can be expressed as
    \begin{equation}
        \rkqm{\left(\H \left[\begin{array}{c}
        \B \\ \hline
        \b
        \end{array}\right]^\top\right)} = \errWeight + 1 \quad \forall\,\b \in \Fq^{\n} \setminus\SumRankSupp{(\E)} \;\;\text{s.t.}\;\SumRankWeight{\n}{(\b)} = 1
    \end{equation}
    where $\B$ is a basis of the row support of the error with respect to the sum-rank metric as in~\eqref{eq:def_B}.
\end{enumerate}
For $t\leq \dminsr - 2$, the second condition is always true, which can be shown by applying~\cite[Lemma 8]{puchingerGenericDecodingSumRank2022}.
However, for $t\geq \dminsr -1$, the decoder becomes probabilistic and returns a unique solution to the decoding problem only if the second condition is satisfied. When considering the average over all error matrices $\E$, the probability of this condition being met becomes a property of the code itself, as it depends on the parity-check matrix $\H$ and therefore on the code's distance spectrum.
Note that the decoder in Algorithm~\ref{alg:decode_high_order_int_sum_rank} can correct errors with a maximum weight of $\errWeight \leq \min\{ n-k-1, \mu\shots \}$. The term $n-k-1$ ensures that the common parity-check matrix of the error code and the component code has at least one non-zero row, which is necessary for successful decoding. The term $\mu\shots$ represents the maximum sum-rank weight for the given parameters, as defined in~\eqref{eq:defsumrankweight}, with $\mu$ given in~\eqref{eq:defmu}.

\begin{figure}[htb]
    \centering
    \begin{tikzpicture}[
        node distance=1cm,
        every node/.style={
            rectangle,
            rounded corners,
            draw=black,
            thick,
            text width=3cm,
            minimum height=0.6cm,
            text centered,
            font=\scriptsize
        },
        arrow/.style={
            thick,
            ->,
            >=stealth
        }
    ]

    \node (s_t_large) {$\intOrder-\errWeight$ is large};
    \node (m_large_1) [right=1cm of s_t_large] {$m$ is large};
    \node (full_rank) [below=1cm of s_t_large, xshift=2cm] {Full-rank condition satisfied with high probability};
    \node (unique_decoding) [below=of full_rank] {Unique decoding for $\errWeight \leq \dminsr-2$};
    \node (high_prob_decoding) [right=of unique_decoding] {Decoding with high probability for $\errWeight \leq n-k-1$};

    \draw[arrow] (s_t_large) -- (full_rank) node [midway, above, sloped, draw=none, xshift=-5pt] {};
    \draw[arrow] (m_large_1) -- (full_rank) node [midway, above, sloped, draw=none, xshift=5pt] {};
    \draw[arrow] (full_rank) -- (unique_decoding) node [midway, right, draw=none] {};
    \draw[arrow] (m_large_1) -- (high_prob_decoding) node [midway, right, draw=none] {};

    \end{tikzpicture}
    \caption{Relationships between parameters, conditions and decoding success for uniform errors.}
    \label{fig:decoding_success}
\end{figure}
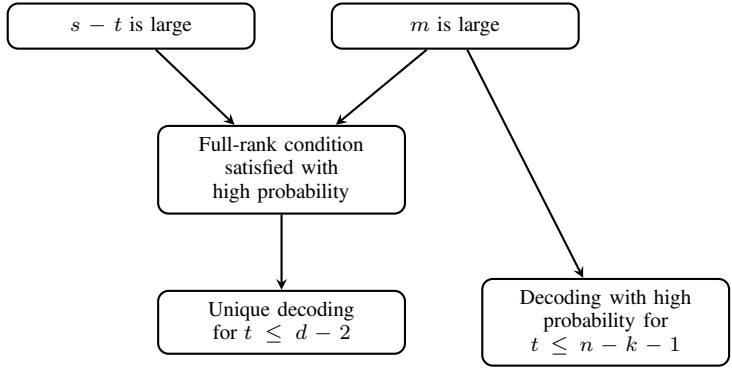

Figure~\ref{fig:dec_rad_illu} illustrates the decoding regions for Algorithm~\ref{alg:decode_high_order_int_sum_rank} when the error matrix $\E$ satisfies the full-rank condition, i.e., $\rkqm(\E)=\errWeight$. This condition is crucial for the success of the decoding algorithm. Figure~\ref{fig:decoding_success} further explores the relationships between various conditions and the decoding success for error matrices drawn uniformly at random. It shows that when the conditions as in Theorem~\ref{thm:failure_bound} are met, such as $\intOrder-\errWeight$ or $m$ being large, the probability of the full-rank condition being satisfied is high. Consequently, this leads to two important results: (1) unique decoding is always possible for $\errWeight \leq \dminsr-2$ when the full-rank condition is satisfied, and (2) decoding is possible with high probability for $\errWeight \leq n-k-1$ when $m$ is large.
\begin{figure}[htp]
    \centering
    \begin{tikzpicture}[
    axis/.style={thick, -{Latex[length=3mm]}, line width=1.0pt}, %
    region1/.style={fill=blue!20, draw=none}, %
    region2/.style={pattern=north east lines, pattern color=gray!40, draw=none}, %
    label/.style={anchor=north, align=center, font=\scriptsize}
]

\coordinate (O) at (0,0);
\coordinate (A) at (4,0);
\coordinate (B) at (8,0);

\draw (O) -- ++(0,-0.1) -- ++(0,0.2);
\draw (A) -- ++(0,-0.1) -- ++(0,0.2);
\draw (B) -- ++(0,-0.1) -- ++(0,0.2);

\node[below] at (O) {$0$};
\node[below] at (A) {$d-2$};
\node[below] at (B) {$\min(\mu\shots, n-k-1)$};

\draw[region1] (O) rectangle (A |- {0,1.1}) node[label, above] at (2,-0) {Unique decoding\\always possible}; %
\draw[region2] (A) rectangle (B |- {0,1.1}) node[label, above] at (6,-0) {Probabilistic\\decoding}; %

\draw[axis] (-0.5,0) -- (9.5,0) node[right] {$t$};

\end{tikzpicture}
    \caption{Illustration of the decoding regions for Algorithm~\ref{alg:decode_high_order_int_sum_rank} if full-rank condition is satisfied.}
    \label{fig:dec_rad_illu}
\end{figure}
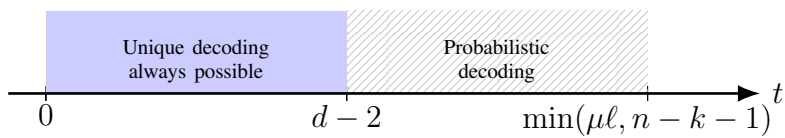

\begin{theorem}\label{thm:success_probability_bound}
    Let $\H \in \Fqm^{(n-k)\times n}$ be a matrix chosen uniformly at random from $\Fqm^{(n-k)\times n}$. We assume that $q^m$ is large enough such that the probability of $\H$ having full $\Fqm$-rank is close to 1. Consider an error matrix $\E$ picked uniformly at random from the set $\set{E}_{\errWeight}^{(\n)}$, where $\E$, $\A$, and $\B$ are as in \eqref{eq:error_decomp}, and $\SumRankWeight{\n}(\E) = \errWeight = \sum_{i=1}^{\ell} \errWeight_i$, satisfying the full-rank condition, i.e., $\rkqm(\E) = \errWeight$. Then, on average, the probability that the condition \eqref{eq:rankBandb} is satisfied is bounded from below and above as follows
    \begin{equation}
        \PLB \leq \Pr[\text{\eqref{eq:rankBandb} is satisfied}] \leq \PUB
    \end{equation}
    where
    \begin{equation}\label{eq:success_probability_bound_lower}
        \PLB \defeq \left(1-\frac{1}{|\set{E}_{\errWeight}^{(\n)}|}\cdot\sum_{\errWeightVec\in\wdecomp{\errWeight,\shots,\mu}}\prod_{i=1}^{\shots}\NM_q(\intOrder m,\eta,\errWeight_i) \cdot \frac{N_{\errWeightVec}}{q^{m(n-k-t)}}\right) \cdot  \prod_{j=0}^{\errWeight-1} \left(1-\frac{1}{q^{m(n-k-j)}}\right)
    \end{equation}
    with
    \begin{equation}\label{eq:success_probability_bound_upper}
        N_{\errWeightVec} \defeq \min\{q^{m(n-k)},\sum_{i=1}^{\shots} (q^{n_i} - q^{\errWeight_i}) \}
    \end{equation}
    and
    \begin{equation}
        \PUB \defeq \prod_{j=0}^{\errWeight} \left(1-\frac{1}{q^{m(n-k-j)}}\right).
    \end{equation}
\end{theorem}

\begin{IEEEproof}
    The proof consists of two parts, one for the lower bound and one for the upper bound.

    First, we show the lower bound. Condition \eqref{eq:rankBandb} can only be satisfied if $\H \B^\top$ is of full $\Fqm$-rank. Since $\H$ is chosen uniformly at random, $\H \B^\top$ is also a matrix uniformly distributed over $\Fqm^{(n-k) \times \errWeight}$. The probability of $\H \B^\top$ having full $\Fqm$-rank is given by
    \begin{equation}
        \pFullRankB \defeq \prod_{j=0}^{\errWeight-1} \left(1-\frac{1}{q^{m(n-k-j)}}\right).
    \end{equation}
    Now, consider a specific vector $\b\in\Fq^{\n} \setminus\SumRankSupp{(\E)}$ and append it to $\B$. Note that for the bound we omit the restriction with $\SumRankWeight{\n}{(\b)}=1$. The probability that $\H [\B^\top \mid \b^\top]$ is of full $\Fqm$-rank, given that $\H \B^\top$ is of full $\Fqm$-rank, is equal to
    \begin{equation}
        \pFullRankBb \defeq \left(1-\frac{1}{q^{m(n-k-\errWeight)}}\right)
    \end{equation}
    which is the probability that the $(\errWeight+1)$-th additional column in $\H [\B^\top \mid \b^\top]$ is linearly independent of the $\errWeight$ remaining columns.
    This must hold true for any $\b\in\Fq^{\n} \setminus\SumRankSupp{(\E)}$ simultaneously.
    Define the event $\mathcal{Z}_i$ as the $(\errWeight+1)$-th column in $\H [\B^\top \mid \b_i^\top]$ for a given $\b_i\in\Fq^{\n} \setminus\SumRankSupp{(\E)}$ being linearly dependent on the remaining $\errWeight$ columns, with $i\in\{1,\ldots,N_{\errWeightVec}\}$ and
    \begin{equation}
        N_{\errWeightVec}  \geq \min\{q^{m(n-k)}, |\Fq^{\n} \setminus\SumRankSupp{(\E)}|\}.
    \end{equation}
    The cardinality $|\Fq^{\n} \setminus\SumRankSupp{(\E)}|$ is given by
    \begin{equation}
        \sum_{i=1}^{\shots}(q^{n_i} - q^{t_i}),
    \end{equation}
    which is the sum of the cardinalities of the blocks that correspond to $|\Fq^{n_i} \setminus \Rowspace{\B^{(i)}}| = q^{n_i} - q^{t_i}$.
    
    By applying the union bound on the events $\mathcal{Z}_i$, the probability that \eqref{eq:rankBandb} is not satisfied is bounded from above as
    \begin{align}
        \Pr[\text{\eqref{eq:rankBandb} is not satisfied}] &\leq (1-\pFullRankB) + \pFullRankB \cdot \sum_{\errWeightVec\in\wdecomp{t,\shots,\mu}}\Pr[\errWeightVec] \cdot \Pr\left[\bigcup_{i=1}^{N_{\errWeightVec}}\mathcal{Z}_i\right]  \\
        &\leq (1-\pFullRankB) + \pFullRankB \cdot \sum_{\errWeightVec\in\wdecomp{t,\shots,\mu}}\Pr[\errWeightVec] \cdot \sum_{i=1}^{N_{\errWeightVec}} \Pr[\mathcal{Z}_i]
    \end{align}
    where $\Pr[\errWeightVec]$ is the marginal distribution of the rank profiles, given by
    \begin{equation}
        \Pr[\errWeightVec] = \frac{1}{|\set{E}_\errWeight^{(\n)}|} \prod_{i=1}^{\shots}\NM_q(\intOrder m,\eta,\errWeight_i).
    \end{equation}
    Now, assuming that all $\mathcal{Z}_i$ are independent, we have that for a given $\errWeightVec$,
    \begin{equation}
        \Pr[\mathcal{Z}_i] = 1-\pFullRankBb = \frac{1}{q^{m(n-k-\errWeight)}}.
    \end{equation}
    Therefore,
    \begin{align}
        \Pr[\text{\eqref{eq:rankBandb} is not satisfied}] &\leq (1-\pFullRankB) + \pFullRankB \cdot \sum_{\errWeightVec\in\wdecomp{t,\shots,\mu}}\Pr[\errWeightVec] \cdot \sum_{i=1}^{N_{\errWeightVec}} \Pr[\mathcal{Z}_i] \\
        &= (1-\pFullRankB) + \pFullRankB \cdot \sum_{\errWeightVec\in\wdecomp{t,\shots,\mu}}\frac{1}{|\set{E}_\errWeight^{(\n)}|} \prod_{i=1}^{\shots}\NM_q(\intOrder m,\eta,\errWeight_i) \cdot \sum_{i=1}^{N_{\errWeightVec}} \frac{1}{q^{m(n-k-\errWeight)}} \\
        &= (1-\pFullRankB) + \pFullRankB \cdot \sum_{\errWeightVec\in\wdecomp{t,\shots,\mu}}\frac{1}{|\set{E}_\errWeight^{(\n)}|} \prod_{i=1}^{\shots}\NM_q(\intOrder m,\eta,\errWeight_i) \cdot \frac{N_{\errWeightVec}}{q^{m(n-k-\errWeight)}}.
    \end{align}
    Consequently,
    \begin{align}
        \Pr[\text{\eqref{eq:rankBandb} is satisfied}] &= 1 - \Pr[\text{\eqref{eq:rankBandb} is not satisfied}] \\
        &\geq 1 - \left((1-\pFullRankB) + \pFullRankB \cdot \sum_{\errWeightVec\in\wdecomp{t,\shots,\mu}} \frac{1}{|\set{E}_{\errWeight}^{(\n)}|} \prod_{i=1}^{\shots}\NM_q(\intOrder m,\eta,\errWeight_i) \cdot \frac{N_{\errWeightVec}}{q^{m(n-k-\errWeight)}}\right) \\
        &= \pFullRankB - \pFullRankB \cdot \sum_{\errWeightVec\in\wdecomp{t,\shots,\mu}} \frac{N_{\errWeightVec} \cdot \prod_{i=1}^{\shots}\NM_q(\intOrder m,\eta,\errWeight_i)}{|\set{E}_{\errWeight}^{(\n)}| \cdot q^{m(n-k-\errWeight)}} \\
        &= \pFullRankB \cdot \left(1 - \sum_{\errWeightVec\in\wdecomp{t,\shots,\mu}} \frac{N_{\errWeightVec} \cdot \prod_{i=1}^{\shots}\NM_q(\intOrder m,\eta,\errWeight_i)}{|\set{E}_{\errWeight}^{(\n)}| \cdot q^{m(n-k-\errWeight)}}\right) \\
        &= \left(1-\sum_{\errWeightVec\in\wdecomp{\errWeight,\shots,\mu}} \frac{N_{\errWeightVec} \cdot \prod_{i=1}^{\shots}\NM_q(\intOrder m,\eta,\errWeight_i)}{|\set{E}_{\errWeight}^{(\n)}| \cdot q^{m(n-k-\errWeight)}}\right)  \prod_{j=0}^{\errWeight-1} \left(1-\frac{1}{q^{m(n-k-j)}}\right)
    \end{align}
    which establishes the lower bound $\PLB$.

    For the upper bound, we observe that the probability that condition \eqref{eq:rankBandb} is satisfied is upper bounded by the event that at least one matrix $\H [\B^\top \mid \b^\top] \in \Fqm^{(n-k) \times (\errWeight+1)}$ is of full $\Fqm$-rank, where $\b\in\Fq^{\n} \setminus\SumRankSupp{(\E)}$. This probability is equal to the probability that a random matrix in $\Fqm^{(n-k) \times (\errWeight+1)}$ is of full $\Fqm$-rank~(see~\cite{lidl_finite_1996}), which is given by
    \begin{equation}
        \PUB \defeq \prod_{j=0}^{\errWeight} \left(1-\frac{1}{q^{m(n-k-j)}}\right).
    \end{equation}
    This completes the proof.
\end{IEEEproof}

\subsection{Simulation Results}

We now investigate the tightness of the upper and lower bounds on the failure probability of condition~\eqref{eq:rankBandb} derived in Theorem~\ref{thm:success_probability_bound}. While these bounds provide theoretical guarantees, they may not always give a precise estimate of the actual failure probability. To assess their accuracy and explore alternative approximations, we conduct simulations.

The following presents an approximation obtained by modifying the proof of Theorem~\ref{thm:success_probability_bound}. Although this approximation does not provide strict bounds, it may yield more realistic estimates of the failure probability and serves as a basis for comparison with the simulated results.

    If, in the proof of Theorem~\ref{thm:success_probability_bound}, we ignore the dependence of the events $\mathcal{Z}_i$ for $i\inshotsarg{N_{\errWeightVec}}$, we obtain neither a lower nor an upper bound on the failure/success probability of condition~\eqref{eq:rankBandb} for a random parity-check matrix $\H$. Nevertheless, we state the expression under that circumstance and use it as an approximation. We then show through simulation that this approximation provides a more realistic estimate of the success probability for relative small $\shotLength$. From the proof of Theorem~\ref{thm:success_probability_bound}, it is straightforward to show that, in this case,
    \begin{equation}\label{eq:success_approx}
        \Pr[\text{condition \eqref{eq:rankBandb} is satisfied}] \approx \prod_{j=0}^{\errWeight-1} \left(1-\frac{1}{q^{m(n-k-j)}}\right) \cdot \sum_{\errWeightVec\in\wdecomp{\errWeight,\shots,\mu}}\Pr[\errWeightVec]\left(1-\frac{1}{q^{m(n-k-\errWeight)}}\right)^{N_{\errWeightVec}}.
    \end{equation}
    
It is worth noting that, in the case of the Hamming metric, the events $\mathcal{Z}_i$ for $i \inshotsarg{N_{\errWeightVec}}$ are actually independent. This is because the rows of the matrix $\B$ consist solely of (scaled) unit vectors. For the Hamming metric, we have $N_{\errWeightVec} = n-t$. When multiplying $\B^\top$ on the right side of $\H$, we effectively select specific columns of $\H$. Moreover, for any additional unit vector $\b$, we select another column from $\H$ to form $\H \cdot [\B^\top \mid \b^\top]$, choosing from the remaining $n-t$ columns. Given the assumption that the entries of $\H$ are independently and uniformly distributed, these selected columns are also independent.

In contrast, this independence does not hold for the rank metric. In the rank-metric case, the matrices $\B^\top$ and $[\B^\top \mid \b^\top]$ can be any full-rank matrices, rather than being limited to (scaled) unit vectors. Consequently, when multiplying these matrices on the right side of $\H$, we obtain linear combinations of the columns of $\H$ rather than simply selecting individual columns. These linear combinations introduce dependencies among the events $\mathcal{Z}_i$, violating the independence assumption.

We investigate the tightness of the upper and lower bounds on the failure probability of condition~\eqref{eq:rankBandb} derived in Theorem~\ref{thm:success_probability_bound} by comparing them with simulated values and an approximation (given in~\eqref{eq:success_approx}). The simulation was performed using a Monte Carlo approach with $10^5$ samples for each point. Each sample involved picking a random parity-check matrix and evaluating the failure probability. We have implemented the simulation with the help of the computer-algebra system SageMath~\cite{sagemath}.

\begin{figure}[htbp]
  \centering

  \newcommand{\ploteta}{1}
  \newcommand{\plotell}{24}
  \newcommand{\plotn}{24}
  \newcommand{\plotq}{2}
  \newcommand{\plotm}{2}
  \newcommand{\plotk}{8}

  \newcommand{\datafile}{./data/sim_dec_rad_vs_t_eta1_ell24_n24_q2_m2_k8.txt}
  
  \begin{tikzpicture}
    \begin{axis}[
        width=\linewidth, %
        height=8cm,
        grid=major, 
        grid style={dotted,gray!80},
        xlabel={$\errWeight$}, %
        ylabel={Success Probability},
        xmin=10,
        xmax=15.0,
        ymin=0,
        ymax=1,
        xtick=data,
        tick label style={font=\scriptsize},
        legend style={font=\scriptsize, at={(0.03,0.03)}, anchor=south west}, 
        label style={inner sep=0, font=\small},
      ]
      
      \addplot[simulated] 
      table[x=t,y=ps,col sep=comma] {\datafile}; 
      \addplot[lowerbound] 
      table[x=t,y=sl,col sep=comma] {\datafile}; 
      \addplot[upperbound] 
      table[x=t,y=su,col sep=comma] {\datafile}; 
      \addplot[approximation] 
      table[x=t,y=sa,col sep=comma] {\datafile}; 

      \legend{Simulated, Lower Bound~\eqref{eq:success_probability_bound_lower}, Upper Bound~\eqref{eq:success_probability_bound_upper}, Approximation~\eqref{eq:success_approx}}    
    \end{axis}
  \end{tikzpicture}
  \caption{Success probability vs error weight $\errWeight$ for $q=\plotq$, $m=\plotm$, $n=\plotn$, $k=\plotk$, $\eta=\ploteta$, and $\ell=\plotell$ with interleaving order $\intOrder = \errWeight$.}
  \label{fig:sim_dec_rad_vs_errWeight_1}
\end{figure}
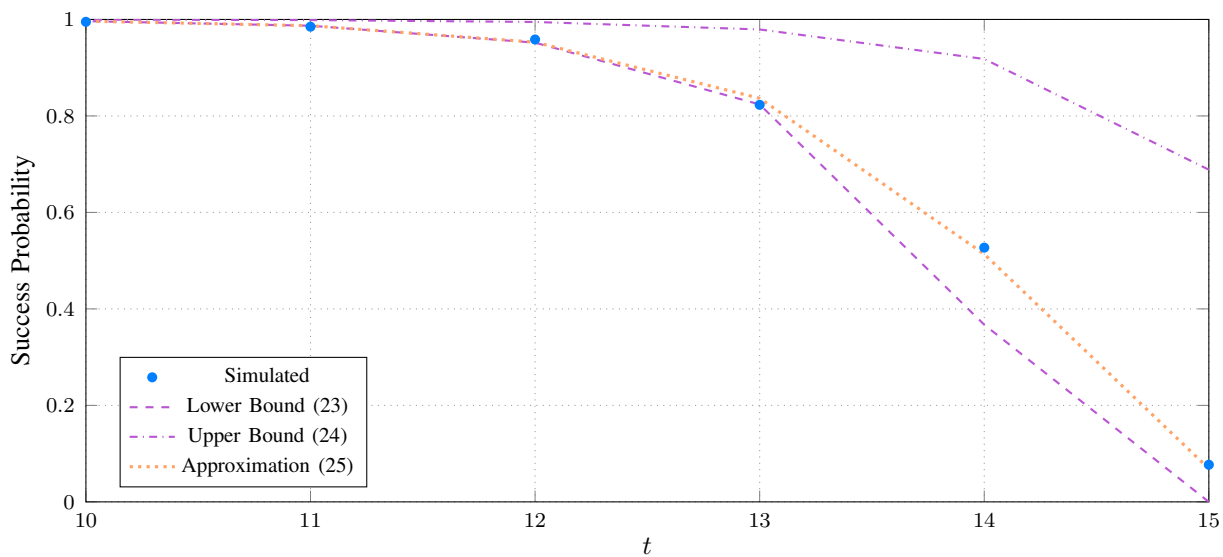

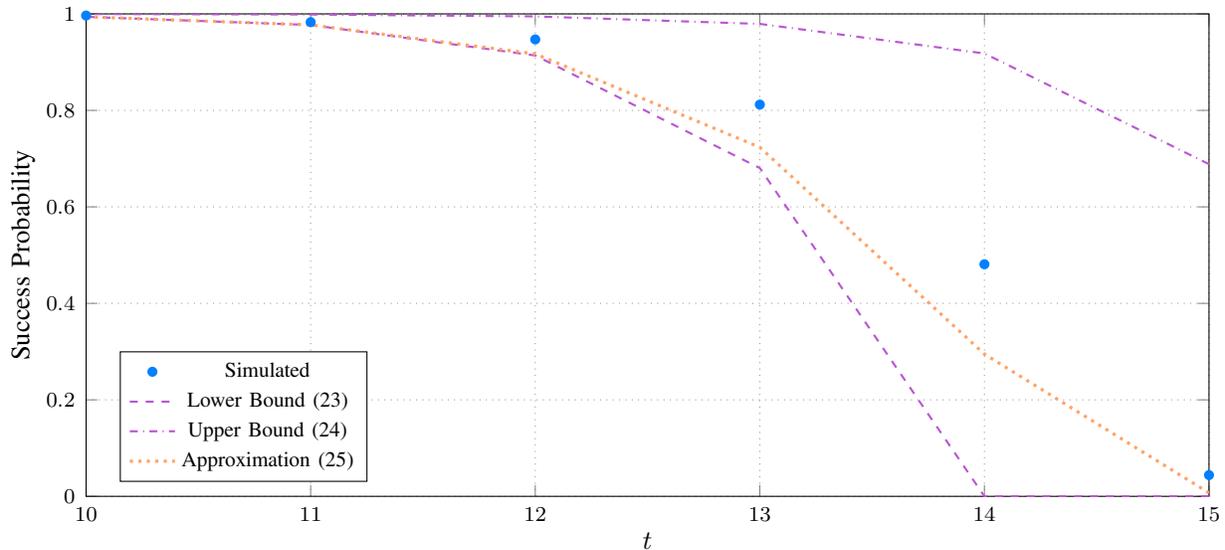
\begin{figure}[htbp]
  \centering

  \newcommand{\ploteta}{2}
  \newcommand{\plotell}{12}
  \newcommand{\plotn}{24}
  \newcommand{\plotq}{2}
  \newcommand{\plotm}{2}
  \newcommand{\plotk}{8}

  \newcommand{\datafile}{./data/sim_dec_rad_vs_t_eta\ploteta_ell\plotell_n\plotn_q\plotq_m\plotm_k\plotk.txt}
  
  \begin{tikzpicture}
    \begin{axis}[
        width=\linewidth, %
        height=8cm,
        grid=major, 
        grid style={dotted,gray!80},
        xlabel={$\errWeight$}, %
        ylabel={Success Probability},
        xmin=10,
        xmax=15.0,
        ymin=0,
        ymax=1,
        xtick=data,
        tick label style={font=\scriptsize},
        legend style={font=\scriptsize, at={(0.03,0.03)}, anchor=south west}, 
        label style={inner sep=0, font=\small},
      ]
      
      \addplot[simulated] 
      table[x=t,y=ps,col sep=comma] {\datafile}; 
      \addplot[lowerbound] 
      table[x=t,y=sl,col sep=comma] {\datafile}; 
      \addplot[upperbound] 
      table[x=t,y=su,col sep=comma] {\datafile}; 
      \addplot[approximation] 
      table[x=t,y=sa,col sep=comma] {\datafile}; 

      \legend{Simulated, Lower Bound~\eqref{eq:success_probability_bound_lower}, Upper Bound~\eqref{eq:success_probability_bound_upper}, Approximation~\eqref{eq:success_approx}}    
    \end{axis}
  \end{tikzpicture}
  \caption{Success probability vs error weight $\errWeight$ for $q=\plotq$, $m=\plotm$, $n=\plotn$, $k=\plotk$, $\eta=\ploteta$, and $\ell=\plotell$ with interleaving order $\intOrder = \errWeight$.}
  \label{fig:sim_dec_rad_vs_errWeight_2}
\end{figure}

\begin{figure}[htbp]
  \centering

  \newcommand{\ploteta}{3}
  \newcommand{\plotell}{8}
  \newcommand{\plotn}{24}
  \newcommand{\plotq}{2}
  \newcommand{\plotm}{2}
  \newcommand{\plotk}{8}

  \newcommand{\datafile}{./data/sim_dec_rad_vs_t_eta\ploteta_ell\plotell_n\plotn_q\plotq_m\plotm_k\plotk.txt}
  
  \begin{tikzpicture}
    \begin{axis}[
        width=\linewidth, %
        height=8cm,
        grid=major, 
        grid style={dotted,gray!80},
        xlabel={$\errWeight$}, %
        ylabel={Success Probability},
        xmin=10,
        xmax=15.0,
        ymin=0,
        ymax=1,
        xtick=data,
        tick label style={font=\scriptsize},
        legend style={font=\scriptsize, at={(0.03,0.03)}, anchor=south west}, 
        label style={inner sep=0, font=\small},
      ]
      
      \addplot[simulated] 
      table[x=t,y=ps,col sep=comma] {\datafile}; 
      \addplot[lowerbound] 
      table[x=t,y=sl,col sep=comma] {\datafile}; 
      \addplot[upperbound] 
      table[x=t,y=su,col sep=comma] {\datafile}; 
      \addplot[approximation] 
      table[x=t,y=sa,col sep=comma] {\datafile}; 

      \legend{Simulated, Lower Bound~\eqref{eq:success_probability_bound_lower}, Upper Bound~\eqref{eq:success_probability_bound_upper}, Approximation~\eqref{eq:success_approx}}    
    \end{axis}
  \end{tikzpicture}
  \caption{Success probability vs error weight $\errWeight$ for $q=\plotq$, $m=\plotm$, $n=\plotn$, $k=\plotk$, $\eta=\ploteta$, and $\ell=\plotell$ with interleaving order $\intOrder = \errWeight$.}
  \label{fig:sim_dec_rad_vs_errWeight_3}
\end{figure}
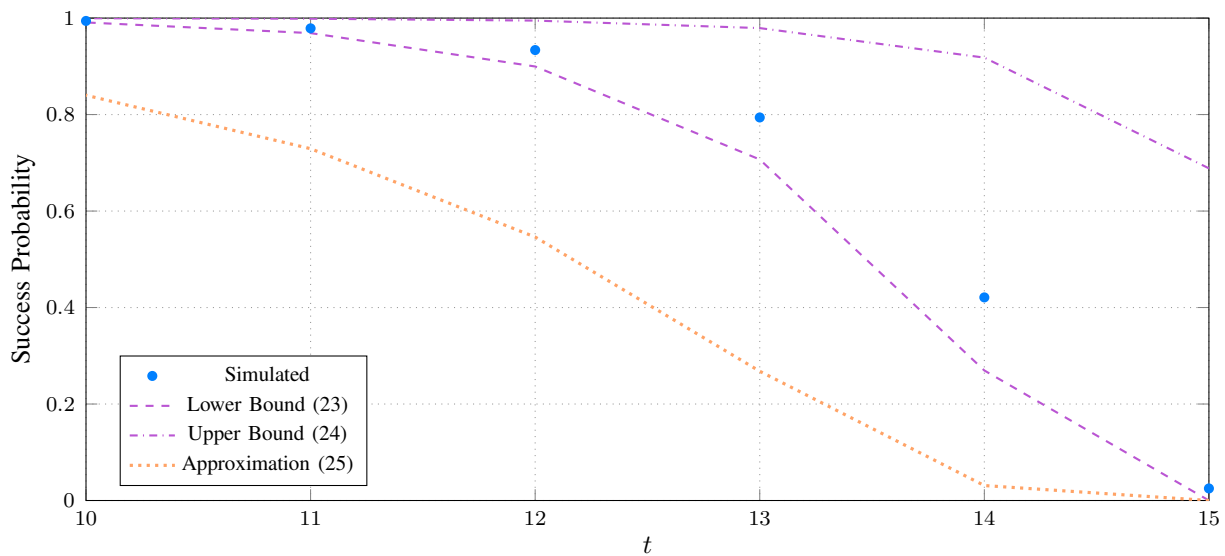

Figure~\ref{fig:sim_dec_rad_vs_errWeight_1} shows parameters $\shotLength = 1$, $\shots = 24$, $n = 24$, $q = 2$, $m = 2$, and $k = 8$, which correspond to the Hamming metric. In this case, we observe that the approximation closely matches the simulated values, and both the upper and lower bounds hold. This reinforces our theory that the approximation is exact in the Hamming-metric case. In Figures~\ref{fig:sim_dec_rad_vs_errWeight_2} and~\ref{fig:sim_dec_rad_vs_errWeight_3}, we increase $\shotLength$ to $2$ and $3$, respectively, while keeping the code parameters constant (i.e., length and dimension). As we move away from the Hamming metric by increasing $\shotLength$, we observe that the approximation becomes less accurate, and the lower bound provides a better estimate. In all plots, the upper bound is relatively loose compared to the lower bound.

Notably, for all scenarios, the success probability for $\errWeight = 14$, which is the second-largest value possible to decode for the given code parameters, stays above $40\%$, which is a relatively high success probability.

\subsection{Examples}
In this section, we present two examples to illustrate the decoding process using Algorithm~\ref{alg:decode_high_order_int_sum_rank} with small code parameters and randomly chosen codes. The first example demonstrates a successful decoding, while the second example showcases a decoding failure where the condition in~\eqref{eq:rankBandb} is not satisfied.

\begin{example}[Successful Decoding]
Let $\Fqm = \F_{2^3}$ with primitive element $\alpha$ and primitive polynomial $\alpha^3 + \alpha + 1$. Consider an interleaved sum-rank-metric code $\mycode{IC}[\intOrder;\n,k,\dminsr]$ of length $n=6$ with $\n = [2, 2, 2]$, $k=2$, $\eta=2$, $\shots=3$, $d=3$, and $\intOrder=3$, 
defined by
the parity-check matrix
\begin{equation}
    \H = \begin{bmatrix}
        \begin{array}{cc|cc|cc}
            1 & 0 & 0 & 0 & \alpha^2 + 1 & \alpha \\
            0 & 1 & 0 & 0 & 1 & \alpha^2 \\
            0 & 0 & 1 & 0 & \alpha & \alpha \\
            0 & 0 & 0 & 1 & \alpha^2 + \alpha + 1 & \alpha + 1 \\
        \end{array}
    \end{bmatrix}.
\end{equation}

Suppose the codeword
\begin{equation}
    \C = \begin{bmatrix}
        \begin{array}{cc|cc|cc}
            \alpha^2 + 1 & 1 & 1 & 1 & \alpha^2 + \alpha & \alpha + 1 \\
            \alpha + 1 & \alpha^2 + \alpha & \alpha^2 & \alpha & 1 & \alpha + 1 \\
            0 & 0 & 1 & \alpha & \alpha^2 & 1 \\
        \end{array}
    \end{bmatrix}
\end{equation}
is corrupted by an error
\begin{equation}
    \E = \begin{bmatrix}
        \begin{array}{cc|cc|cc}
            0 & \alpha^2 + 1 & \alpha^2 + 1 & \alpha^2 + 1 & 0 & 0 \\
            1 & 0 & \alpha^2 & \alpha^2 & 0 & 0 \\
            \alpha + 1 & \alpha & \alpha + 1 & \alpha + 1 & 0 & 0 \\
        \end{array}
    \end{bmatrix}
\end{equation}
with $\C\H^\top=\bm{0}$, $\rkqm(\E) = \SumRankWeight{\n}(\E) = 3$ and $\errWeightVec = [2,1,0]$. The received word is $\Y = \C + \E$, given by
\begin{equation}
    \Y = \begin{bmatrix}
        \begin{array}{cc|cc|cc}
            \alpha^2 + 1 & \alpha^2 & \alpha^2 & \alpha^2 & \alpha^2 + \alpha & \alpha + 1 \\
            \alpha & \alpha^2 + \alpha & 0 & \alpha^2 + \alpha & 1 & \alpha + 1 \\
            \alpha + 1 & \alpha & \alpha & 1 & \alpha^2 & 1 \\
        \end{array}
    \end{bmatrix}.
\end{equation}

The syndrome $\S = \H\Y^\top$ is computed as
\begin{equation}
    \S = \begin{bmatrix}
        0 & 1 & \alpha + 1 \\
        \alpha^2 + 1 & 0 & \alpha \\
        \alpha^2 + 1 & \alpha^2 & \alpha + 1 \\
        \alpha^2 + 1 & \alpha^2 & \alpha + 1 \\
    \end{bmatrix}.
\end{equation}

We find a matrix $\P\in\Fqm^{(n-k)\times (n-k)}$ with $\rkqm(\P) = n-k = 4$, such as
\begin{equation}
    \P = \begin{bmatrix}
        1 & \alpha^2 + 1 & 0 & \alpha^2 + \alpha + 1 \\
        \alpha + 1 & \alpha^2 + 1 & 0 & \alpha^2 + 1 \\
        \alpha^2 + \alpha + 1 & \alpha + 1 & 0 & \alpha + 1 \\
        0 & 0 & 1 & 1 \\
    \end{bmatrix},
\end{equation}
which transforms $\P\S$ into row-echelon form.

The last $n-k-t=1$ rows of $\P\H$ yield
\begin{equation}
    \Hsub = \begin{bmatrix}
        \begin{array}{cc|cc|cc}
            0 & 0 & 1 & 1 & \alpha^2 + 1 & 1 \\
        \end{array}
    \end{bmatrix}.
\end{equation}

Expanding each sub-block of $\Hsub$ over $\F_{2}$ yields
\begin{equation}
    \ext\left(\Hsub^{(1)}\right) = \begin{bmatrix} 0 & 0 \\ 0 & 0 \\ 0 & 0 \end{bmatrix},\quad
    \ext\left(\Hsub^{(2)}\right) = \begin{bmatrix} 1 & 1 \\ 0 & 0 \\ 0 & 0 \end{bmatrix},\quad
    \ext\left(\Hsub^{(3)}\right) = \begin{bmatrix} 1 & 1 \\ 0 & 0 \\ 1 & 0 \end{bmatrix}.
\end{equation}

Note that $\ext\left(\Hsub^{(1)}\right)$ is an all-zero matrix, indicating that this block corresponds to a full-rank error.

Next, we compute a basis for each of the right kernels of $\ext\left(\Hsub^{(1)}\right)$, $\ext\left(\Hsub^{(2)}\right)$, and $\ext\left(\Hsub^{(3)}\right)$ such that
    \begin{equation}
        \ext\left(\Hsub^{(1)}\right){\B^{(1)}}^\top = \bm{0},\quad
        \ext\left(\Hsub^{(2)}\right){\B^{(2)}}^\top = \bm{0},\quad
        \ext\left(\Hsub^{(3)}\right){\B^{(3)}}^\top = \bm{0}
    \end{equation}
    and
    \begin{align}
    \rkq(\B^{(1)}) &= n_1 - \rkq(\Hsub^{(1)}) = 2,\\
    \rkq(\B^{(2)}) &= n_2 - \rkq(\Hsub^{(2)}) = 1,\\
    \rkq(\B^{(3)}) &= n_3 - \rkq(\Hsub^{(3)}) = 0.    
    \end{align}
    This gives us
    \begin{equation}
        \B^{(1)} = \begin{bmatrix} 1 & 0 \\ 0 & 1 \end{bmatrix}\in\Fq^{2\times 2}, \quad
        \B^{(2)} = \begin{bmatrix} 1 & 1 \end{bmatrix}\in\Fq^{1\times 2}, \quad
        \B^{(3)} = \begin{bmatrix} \quad\quad \end{bmatrix}\in\Fq^{0\times 2}.
    \end{equation}
    The matrix $\B$ is then given by
    \begin{equation}
        \B = \diag\left(\B^{(1)},\B^{(2)},\B^{(3)}\right) = \begin{bmatrix}
        \begin{array}{cc|cc|cc}
            1 & 0 & 0 & 0 & 0 & 0 \\
            0 & 1 & 0 & 0 & 0 & 0 \\
            \hline
            0 & 0 & 1 & 1 & 0 & 0 \\
            \hline
        \end{array}
    \end{bmatrix}
    \end{equation}
    Finally, solving for $\A$, i.e.,
    \begin{align}
        \H\B^\top\A^\top &= \S \\
    \begin{bmatrix}
        1 & 0 & 0 & 0 & 0 \\
        0 & 1 & 0 & 0 & 0 \\
        0 & 0 & 1 & 1 & 0 \\
    \end{bmatrix}\A^\top &= \begin{bmatrix}
        0 & 1 & \alpha + 1 \\
        \alpha^2 + 1 & 0 & \alpha \\
        \alpha^2 + 1 & \alpha^2 & \alpha + 1 \\
        \alpha^2 + 1 & \alpha^2 & \alpha + 1 \\
    \end{bmatrix}         
    \end{align} 
    yields  
    \begin{align}
        &\A^\top = \begin{bmatrix}
        0 & \alpha^2 + 1 & \alpha^2 + 1 \\
        1 & 0 & \alpha^2 \\
        \alpha + 1 & \alpha & \alpha + 1 \\
    \end{bmatrix} \\
    \Longrightarrow&\quad \hat{\E} = \A\B = \begin{bmatrix}
    \begin{array}{cc|c|c}
    0 & \alpha^2 + 1 & \alpha^2 + 1 & 0 \\
    1 & 0 & \alpha^2 & 0 \\
    \alpha + 1 & \alpha & \alpha + 1 & 0 \\
    \end{array}
    \end{bmatrix}
    \end{align}
    and $\hat\E = \E$. 
    Note that decoding is possible since $\rkqm(\H[\B^\top\mid \b^\top]) = t+1 = 4$ for all $\b\in \Fq^{\n} \setminus\SumRankSupp{(\E)}$ such that $\SumRankWeight{\n}{(\b)} = 1$.
\end{example}

\begin{example}[Decoding Failure]
    Let $\Fqm = \F_{2^2}$ with primitive element $\alpha$ and minimal polynomial $\alpha^2 + \alpha + 1$. Further let $\mycode{IC}[\intOrder;\n,k,\dminsr]$ be an interleaved sum-rank-metric code of length $n=6$ with $\n = [2, 2, 2]$, $k=2$, $\dminsr=4$, $\eta=2$, $\shots=3$ and $\intOrder=3$, defined by the parity-check matrix
    \begin{equation}
        \H = \begin{bmatrix}
        \begin{array}{cc|cc|cc}
            1 & 0 & 0 & \alpha + 1 & 0 & \alpha \\
            0 & 1 & 0 & 1 & 0 & 1 \\
            0 & 0 & 1 & \alpha & 0 & 0 \\
            0 & 0 & 0 & 0 & 1 & \alpha + 1 \\
        \end{array}
    \end{bmatrix}.
    \end{equation}
    Suppose that the codeword
    \begin{equation}
        \C = \begin{bmatrix}
        \begin{array}{cc|cc|cc}
            0 & \alpha & 1 & \alpha + 1 & \alpha + 1 & 1 \\
            \alpha & 0 & \alpha + 1 & \alpha & 1 & \alpha \\
            0 & \alpha & 1 & \alpha + 1 & \alpha + 1 & 1 \\
        \end{array}
    \end{bmatrix}
    \end{equation}
    is corrupted by an error 
    \begin{equation}
    \E =  \begin{bmatrix}
        \begin{array}{cc|cc|cc}
            \alpha & 0 & \alpha & \alpha & 0 & 0 \\
            1 & 1 & \alpha + 1 & \alpha + 1 & 0 & 0 \\
            \alpha & 1 & \alpha + 1 & \alpha + 1 & 0 & 0 \\
        \end{array}
    \end{bmatrix}
    \end{equation}
    with $\rkqm(\E) = \SumRankWeight{\n}(\E) = 3$ and $\errWeightVec = [2,1,0]$.
    The resulting received word is then $\Y = \C + \E$ and thus
    \begin{equation}
        \Y =     \begin{bmatrix}
        \begin{array}{cc|cc|cc}
            \alpha & \alpha & \alpha + 1 & 1 & \alpha + 1 & 1 \\
            \alpha + 1 & 1 & 0 & 1 & 1 & \alpha \\
            \alpha & \alpha + 1 & \alpha & 0 & \alpha + 1 & 1 \\
        \end{array}
    \end{bmatrix}.
    \end{equation}
    The syndrome is then
    \begin{equation}
        \S = \H\Y^\top = \begin{bmatrix}
        \alpha + 1 & \alpha + 1 & 0 \\
        \alpha & \alpha & \alpha \\
        1 & \alpha & \alpha \\
        0 & 0 & 0 \\
    \end{bmatrix}.
    \end{equation}
    We can find $\P\in\Fqm^{(n-k)\times (n-k)}$ with $\rkqm(\P) = n-k = 4$, hence
    \begin{equation}
        \P = \begin{bmatrix}
        0 & \alpha & \alpha & 0 \\
        \alpha & \alpha & \alpha & 0 \\
        \alpha & \alpha + 1 & 0 & 0 \\
        0 & 0 & 0 & 1 \\
    \end{bmatrix}
    \end{equation}
    such that $\P\S$ is in row-echelon form.
    The last $n-k-t=1$ rows of
    \begin{equation}
        \P\H =     \begin{bmatrix}
        \begin{array}{cc|cc|cc}
            0 & \alpha & \alpha & 1 & 0 & \alpha \\
            \alpha & \alpha & \alpha & 0 & 0 & 1 \\
            \alpha & \alpha + 1 & 0 & \alpha & 0 & 0 \\
            0 & 0 & 0 & 0 & 1 & \alpha + 1 \\
        \end{array}
    \end{bmatrix}
    \end{equation}
    yields
    \begin{equation}
        \Hsub =  \begin{bmatrix}
        \begin{array}{cc|cc|cc}
            0 & 0 & 0 & 0 & 1 & \alpha + 1 \\
        \end{array}
    \end{bmatrix}.
    \end{equation}
    Next we expand every sub-block of $\Hsub$ over $\F_{2}$ and obtain
    \begin{equation}
        \ext\left(\Hsub^{(1)}\right) = \begin{bmatrix} 0 & 0 \\ 0 & 0  \end{bmatrix},\quad
        \ext\left(\Hsub^{(2)}\right) = \begin{bmatrix} 0 & 0 \\ 0 & 0  \end{bmatrix},\quad
        \ext\left(\Hsub^{(3)}\right) = \begin{bmatrix} 1 & 1 \\ 0 & 1  \end{bmatrix}.
    \end{equation}
    Next we compute a basis for each of the right kernels of $\ext\left(\Hsub^{(1)}\right)$, $\ext\left(\Hsub^{(2)}\right)$ and $\ext\left(\Hsub^{(3)}\right)$ such that
    \begin{equation}
        \ext\left(\Hsub^{(1)}\right){\B^{(1)}}^\top = \bm{0},\quad
        \ext\left(\Hsub^{(2)}\right){\B^{(2)}}^\top = \bm{0},\quad
        \ext\left(\Hsub^{(3)}\right){\B^{(3)}}^\top = \bm{0}
    \end{equation}
    and
    \begin{align}
    \rkq(\B^{(1)}) = n_1 - \rkq(\Hsub^{(1)}) = 2,\\
    \rkq(\B^{(2)}) = n_2 - \rkq(\Hsub^{(2)}) = 2,\\
    \rkq(\B^{(3)}) = n_3 - \rkq(\Hsub^{(3)}) = 1,
    \end{align}
    which gives us
    \begin{equation}
        \B^{(1)} = \begin{bmatrix} 1 & 0 \\ 0 & 1 \end{bmatrix}, \quad
        \B^{(2)} = \begin{bmatrix} 1 & 0 \\ 0 & 1 \end{bmatrix}, \quad
        \B^{(3)} = \begin{bmatrix} \quad\quad \end{bmatrix}.
    \end{equation}
    The matrix $\B$ is then given by
    \begin{equation}
        \B = \diag\left(\B^{(1)},\B^{(2)},\B^{(3)}\right) = \begin{bmatrix}
        \begin{array}{cc|cc|cc}
            1 & 0 & 0 & 0 & 0 & 0 \\
            0 & 1 & 0 & 0 & 0 & 0 \\
            \hline
            0 & 0 & 1 & 0 & 0 & 0 \\
            0 & 0 & 0 & 1 & 0 & 0 \\
            \hline
        \end{array}
    \end{bmatrix}.
    \end{equation}
    In fact, we have that $\rkq(\B^{(1)}) + \rkq(\B^{(2)}) + \rkq(\B^{(3)}) = 4 > t = 3$, and therefore we cannot uniquely recover the error $\E$ anymore. This is because the decoding condition in~\eqref{eq:rankBandb} is not satisfied, since there exists $\b\in \Fq^{\n} \setminus\SumRankSupp{(\E)}$ such that $\SumRankWeight{\n}{(\b)} = 1$ and $\rkqm(\H[\B^\top\mid \b^\top]) \neq t+1 = 4$. That is, for $\b = \begin{bmatrix}\begin{array}{cc|cc|cc} 0 & 0 & 1 & 0 & 0 & 0 \end{array}\end{bmatrix}$, we have
    \begin{equation}
        \H \left[\B^\top\mid\b^\top\right] = \begin{bmatrix}
        1 & 0 & \alpha + 1 & 0 \\
        1 & 1 & 1 & 0 \\
        0 & 0 & \alpha + 1 & 1 \\
        0 & 0 & 0 & 0 \\
    \end{bmatrix} \quad\Longrightarrow\quad \rkqm\left(\H \left[\B^\top\mid\b^\top\right]\right) = 3 < 4.
    \end{equation}
\end{example}

\subsection{Special Cases of the Algorithm for Hamming and Rank Metric}

The decoder presented in Algorithm~\ref{alg:decode_high_order_int_sum_rank} is a generalization of the Metzner--Kapturowski decoder for the Hamming metric~\cite{metzner1990general} and the Metzner--Kapturowski-like decoder for the rank metric~\cite{puchinger2019decoding}. In this section, we highlight the differences in how the proposed decoder operates in three distinct metrics: the Hamming metric, the rank metric, and the sum-rank metric. Note that both the Hamming and rank metrics are special cases of the sum-rank metric. We also emphasize the analogous definitions of the error support for all three cases. To differentiate between the error weights in each metric, we use the following notation: $t_{H}$ for the Hamming metric, $t_{R}$ for the rank metric, and $t_{\Sigma R}$ for the sum-rank metric. 

In the Hamming metric, the support of an error matrix $\E$ is defined as the set of indices corresponding to the non-zero columns of $\E$, that is,
\begin{equation*}
    \HammingSupp(\E) \defeq \{j \,:\, \text{the } j\text{-th column of } \E \text{ is non-zero}\}.
\end{equation*}
However, this classical support notion does \emph{not} directly coincide with the definition of the sum-rank support from~\eqref{eq:sumranksupport}. Nonetheless, there is a one-to-one correspondence between these concepts.
We demonstrate this by first describing $\SumRankSupp(\E)$ and then relating it to $\HammingSupp(\E)$.
Since each of the blocks $\shot{\E}{1}, \dots, \shot{\E}{\shots}$ has length one in the Hamming-metric setting, at most one rank error can occur per block.
Thus, the $i$-th block $\shot{\B}{i}$ in the error decomposition $\E = \A \cdot \diag(\shot{\B}{1}, \dots, \shot{\B}{\shots})$ from~\eqref{eq:error_decomp} has size $\shot{t}{i}_H \times 1$ with $\shot{t}{i}_H \in \{0, 1\}$.
If the $i$-th block for an $i \inshots$ is erroneous, the matrix $\shot{\B}{i}$ contains one nonzero $\Fq$ element, which implies $\Rowspace{\shot{\B}{i}} = \Fq$.
If, on the other hand, the block $\shot{\E}{i}$ is error-free, the matrix $\shot{\B}{i}$ has size $0 \times 1$ and its row space $\Rowspace{\shot{\B}{i}}$ is the trivial vector space $\{\0\} \subseteq \Fq$.
Thus, the sum-rank support $\SumRankSupp(\E) = \Rowspace{\shot{\B}{1}} \times \dots \times \Rowspace{\shot{\B}{\shots}}$ of $\E$ is a Cartesian product containing copies of $\Fq$ and $\{\0\}$ in the respective positions.
This allows us to define a bijection between the sum-rank support and the classical definition of Hamming support given above.
Namely,
\begin{align}
    \HammingSupp(\E) &\mapsto \SumRankSupp(\E) = \bigtimes_{i=1}^{n} \X_i \text{ with } \X_i =
    \begin{cases}
        \Fq & \text{if } i \in \HammingSupp(\E) \\
        \{\0\} & \text{if } i \notin \HammingSupp(\E)
    \end{cases}
\end{align}
maps a subset of the indices $\{1, \dots, n\}$ to the corresponding sum-rank support contained in $\Fq^{\n}$ with $\n=[1,\ldots,1]$.
We stick to $\HammingSupp(\E)$ to explain the consequences for decoding in the Hamming metric in the following.

An error matrix $\E$ with $t_{H}$ errors in the Hamming metric can be factored into $\E=\A\B$, where the rows of $\B$ are (scaled) unit vectors corresponding to the $t_H$ error positions.
Consequently, the support of $\E$ is the union of the supports of the rows $\B_i$ of $\B$ ($\forall i\inshotsarg{t_H}$), i.e.,
\begin{equation*}
    \HammingSupp(\E) = \bigcup\limits_{i=1}^{t_H}\HammingSupp(\B_i).
\end{equation*}

\begin{figure}[ht!]
    \centering
    \begin{tikzpicture}[scale=.5]

    \draw[rounded corners, dashed] (-1, -1.5) rectangle (10, 5) {};
    \node[] () at (-0.5, 4.5) {$\E$};

    \draw[nonerrorblock] (0,0) rectangle (0.5,3);
    \draw[nonerrorblock] (0.5,0) rectangle (1,3);
    \draw[nonerrorblock] (1,0) rectangle (1.5,3);
    \draw[nonerrorblock] (1.5,0) rectangle (2,3);
    \draw[errorblock] (2,0) rectangle (2.5,3);
    \draw[errorblock] (2.5,0) rectangle (3,3);
    \draw[nonerrorblock] (3,0) rectangle (3.5,3);
    \draw[nonerrorblock] (3.5,0) rectangle (4,3);
    \draw[nonerrorblock] (4,0) rectangle (4.5,3);
    \draw[nonerrorblock] (4.5,0) rectangle (5,3);
    \draw[nonerrorblock] (5,0) rectangle (5.5,3);
    \draw[nonerrorblock] (5.5,0) rectangle (6,3);
    \draw[errorblock] (6,0) rectangle (6.5,3);
    \draw[nonerrorblock] (6.5,0) rectangle (7,3);
    \draw[errorblock] (7,0) rectangle (7.5,3);
    \draw[nonerrorblock] (7.5,0) rectangle (8,3);
    \draw[nonerrorblock] (8,0) rectangle (8.5,3);
    \draw[nonerrorblock] (8.5,0) rectangle (9,3);

    \draw[-latex] (2.2,-0.5) -- (2.2,0);
    \draw[-latex] (2.7,-0.5) -- (2.7,0);
    \draw[-latex] (6.2,-0.5) -- (6.2,0);
    \draw[-latex] (7.2,-0.5) -- (7.2,0);
    \node[] () at (5.0, -1.0) {error positions};

    \node[] () at (10.7, 1.5) {$=$};

    \draw[rounded corners, dashed] (11.2, -1.5) rectangle (15.2, 5) {};
    \node[] () at (11.7, 4.5) {$\A$};

    \draw[errorblock] (11.7,0) rectangle (12.45,4);
    \draw[errorblock] (12.45,0) rectangle (13.2,4);
    \draw[errorblock] (13.2,0) rectangle (13.95,4);
    \draw[errorblock] (13.95,0) rectangle (14.7,4);

    \node[] () at (15.7, 1.5) {$\cdot$};

    \draw[rounded corners, dashed] (16.2, -1.5) rectangle (25.7, 5) {};
    \node[] () at (16.7, 4.5) {$\B$};

    \draw[] (16.7,0) rectangle (17.2,3);
    \draw[] (17.2,0) rectangle (17.7,3);
    \draw[] (17.7,0) rectangle (18.2,3);
    \draw[] (18.2,0) rectangle (18.7,3);
    \draw[] (18.7,0) rectangle (19.2,3);
    \draw[] (19.2,0) rectangle (19.7,3);
    \draw[] (19.7,0) rectangle (20.2,3);
    \draw[] (20.2,0) rectangle (20.7,3);
    \draw[] (20.7,0) rectangle (21.2,3);
    \draw[] (21.2,0) rectangle (21.7,3);
    \draw[] (21.7,0) rectangle (22.2,3);
    \draw[] (22.2,0) rectangle (22.7,3);
    \draw[] (22.7,0) rectangle (23.2,3);
    \draw[] (23.2,0) rectangle (23.7,3);
    \draw[] (23.7,0) rectangle (24.2,3);
    \draw[] (24.2,0) rectangle (24.7,3);

    \node[] () at (18, 2.6) {\small{1}};
    \node[] () at (18.5, 1.9) {\small{1}};
    \node[] () at (22, 1.2) {\small{1}};
    \node[] () at (23, 0.4) {\small{1}};

    \draw[-latex] (18,-0.5) -- (18,0);
    \draw[-latex] (18.3,-0.5) -- (18.3,0);
    \draw[-latex] (22,-0.5) -- (22,0);
    \draw[-latex] (23,-0.5) -- (23,0);

    \node[] () at (20.2, -1.0) {error positions};

    \node[] () at (-1.7, -5) {$\Hce=$};
    \draw[errorblock] (0, -3.5) rectangle (0.5, -6.5);
    \draw[errorblock] (0.5, -3.5) rectangle (1, -6.5);
    \draw[errorblock] (1, -3.5) rectangle (1.5, -6.5);
    \draw[errorblock] (1.5, -3.5) rectangle (2, -6.5);
    \draw[nonerrorblock] (2, -3.5) rectangle (2.5, -6.5);
    \draw[nonerrorblock] (2.5, -3.5) rectangle (3, -6.5);
    \draw[errorblock] (3, -3.5) rectangle (3.5, -6.5);
    \draw[errorblock] (3.5, -3.5) rectangle (4, -6.5);
    \draw[errorblock] (4, -3.5) rectangle (4.5, -6.5);
    \draw[errorblock] (4.5, -3.5) rectangle (5, -6.5);
    \draw[errorblock] (5, -3.5) rectangle (5.5, -6.5);
    \draw[errorblock] (5.5, -3.5) rectangle (6, -6.5);
    \draw[nonerrorblock] (6, -3.5) rectangle (6.5, -6.5);
    \draw[errorblock] (6.5, -3.5) rectangle (7, -6.5);
    \draw[nonerrorblock] (7, -3.5) rectangle (7.5, -6.5);
    \draw[errorblock] (7.5, -3.5) rectangle (8, -6.5);
    \draw[errorblock] (8, -3.5) rectangle (8.5, -6.5);
    \draw[errorblock] (8.5, -3.5) rectangle (9, -6.5);

    \draw[-latex] (2.2,-7.5) -- (2.2,-6.5);
    \draw[-latex] (2.7,-7.5) -- (2.7,-6.5);
    \draw[-latex] (6.2,-7.5) -- (6.2,-6.5);
    \draw[-latex] (7.2,-7.5) -- (7.2,-6.5);
    \node[] () at (5.0, -8.0) {all-zero columns in error positions};
    \node[anchor=west] () at (10.0, -5) {$\Rightarrow \HammingSupp(\E) =\bigcup\limits_{i=1}^{t_H}\HammingSupp(\B_i) =  $};
    \node[anchor=west] () at (12, -7.5) {$ =[1:n] \setminus \bigcup\limits_{i=1}^{n-k-t_H}\HammingSupp(\H_{\CECode,i})$};

\end{tikzpicture}
    \caption{Illustration of the error support for the Hamming-metric case with $\E=\A\B\in\Fqm^{\intOrder \times n}$, $\A\in\Fqm^{\intOrder\times t_H}$, $\B\in\Fq^{t_H\times n}$ and $\Hsub\in\Fqm^{(n-k-t_H)\times n}$. $\B_i$ is the $i$-th row of $\B$ and $\H_{\CECode,i}$ the $i$-th row of $\Hsub$.}\label{fig:hamming_figure}
\end{figure}
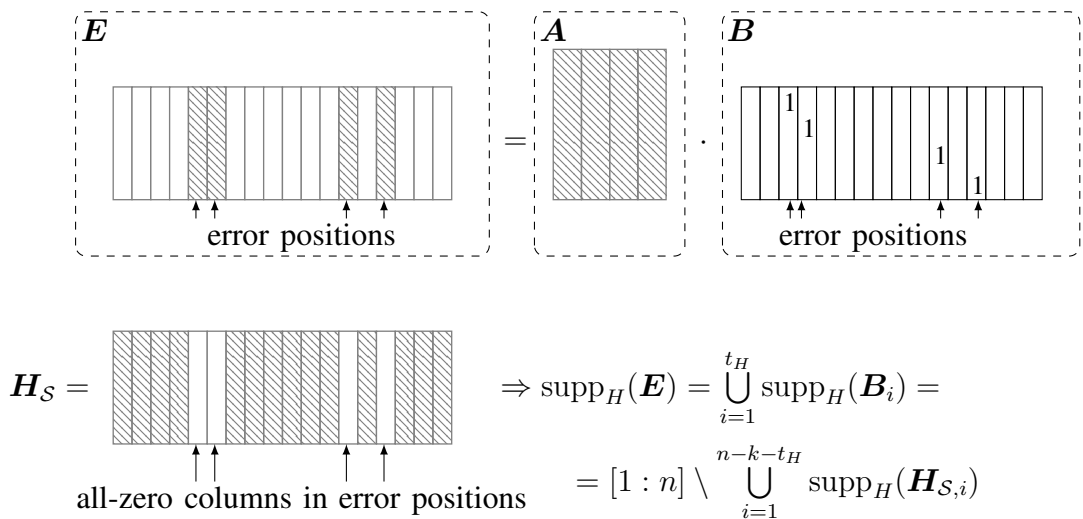

When the full-rank condition for the Metzner--Kapturowski decoder is satisfied, the zero columns in $\Hsub$ reveal the error positions and determine the error support. In this case, we have
\begin{equation*}
    \HammingSupp(\E) = [1:n] \setminus \bigcup\limits_{i=1}^{n-k-t_H} \HammingSupp(\H_{\CECode,i})
\end{equation*}
where $\H_{\CECode,i}$ denotes the $i$-th row of $\Hsub$. Note that this equality corresponds to~\eqref{eq:dualSuppEqualErrorSupp} in the general case.
The process of recovering the error support $\HammingSupp(\E)$ from $\Hsub$ is depicted in Figure~\ref{fig:hamming_figure}.

The rank-metric case is analogous to the Hamming-metric case but with a different definition of the error support. An error matrix $\E$ with rank $\rkq(\E) = t_R$ can be decomposed as $\E = \A\B$. The rank support $\RankSupp(\E)$ of $\E$ is defined as the row space of $\B$, which is spanned by the union of all rows $\B_i$ of $\B$, where $\B_i$ is the $i$-th row of $\B$. This coincides exactly with the more general definition in the sum-rank metric from~\eqref{eq:sumranksupport} for $\shots = 1$.
Thus, the support of $\E$ is given by
\begin{equation*}
    \RankSupp(\E) = \mksum\limits_{i=1}^{t_R} \RankSupp(\B_i)
\end{equation*}
where $\mksum$ denotes the addition of vector spaces, i.e., the span of the union of the considered spaces.
If the full-rank condition on the error matrix is satisfied, the rank support of $\E$ can be determined by the $\Fq$-kernel of $\Hsub$~\cite{renner2021decoding}.The $\Fq$-row space of $\Hsub$ can be computed by taking the span of the union of spaces $\RankSupp(\H_{\CECode,i})$, where $\H_{\CECode,i}$ is the $i$-th row of $\Hsub$. Consequently, the support of $\E$ is given by
\begin{equation*}
    \RankSupp(\E) = \left(\mksum\limits_{j=1}^{n-k-t_R}\RankSupp(\H_{\text{sub},j})\right)^{\perp}.
\end{equation*}

In the sum-rank metric, according to~\eqref{eq:sumranksupport}, we have
\begin{align*}
    \SumRankSupp(\E) &= \RankSupp{(\B^{(1)})} \times \RankSupp{(\B^{(2)})} \times \dots \times \RankSupp{(\B^{(\intOrder)})} \\
    &= \left(\mksum\limits_{j=1}^{n-k-\errWeight_{\Sigma R}} \RankSupp(\B_{j}^{(1)})\right) \times\dots\times\left(\mksum\limits_{j=1}^{n-k-\errWeight_{\Sigma R}} \RankSupp(\B_{j}^{(\shots)})\right).
\end{align*}

Based on Theorem~\ref{thm:errorsupportinFqHdual}, we have
\begin{align*}
    \SumRankSupp(\E) &= \left(\mksum\limits_{j=1}^{n-k-\errWeight_{\Sigma R}} \RankSupp(\H_{\CECode,j}^{(1)})\right)^{\perp} \times\dots\\
    &\dots\times\left(\mksum\limits_{j=1}^{n-k-\errWeight_{\Sigma R}} \RankSupp(\H_{\CECode,j}^{(\intOrder)})\right)^{\perp}.
\end{align*}

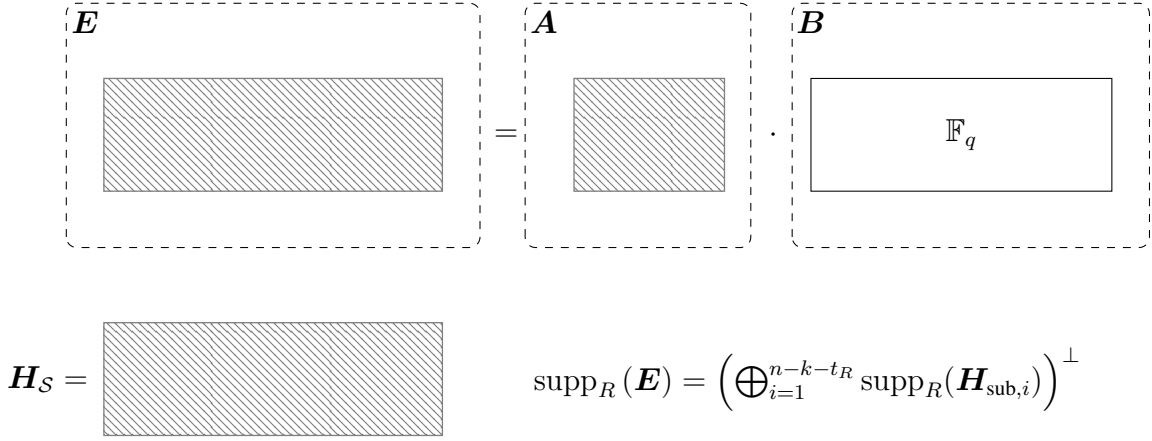
\begin{figure}[ht!]
    \centering
    \begin{tikzpicture}[scale=.5]

    \draw[rounded corners, dashed] (-1, -1.5) rectangle (10, 5) {};
    \node[] () at (-0.5, 4.5) {$\E$};

    \draw[errorblock] (0,0) rectangle (9,3);

    \node[] () at (10.7, 1.5) {$=$};

    \draw[rounded corners, dashed] (11.2, -1.5) rectangle (17.2, 5) {};
    \node[] () at (11.7, 4.5) {$\A$};

    \draw[errorblock] (12.5,0) rectangle (16.5,3);

    \node[] () at (17.8, 1.5) {$\cdot$};

    \draw[rounded corners, dashed] (18.3, -1.5) rectangle (27.8, 5) {};
    \node[] () at (18.8, 4.5) {$\B$};

    \draw[] (18.8,0) rectangle (26.8,3);
    \node[] () at (22.8, 1.5) {$\Fq$};
    
    \node[] () at (-1.5, -5) {$\Hce=$};
    \draw[errorblock] (0, -3.5) rectangle (9, -6.5);

    \node[] () at (18.7, -5) {$\RankSupp{(\E)} = \left(\mksum\limits_{i=1}^{n-k-t_R}\RankSupp(\H_{\text{sub},i})\right)^{\perp}$};

\end{tikzpicture}
    \caption{Illustration of the error support for the rank-metric case with $\E=\A\B\in\Fqm^{\intOrder \times n}$, $\A\in\Fqm^{\intOrder\times t_R}$, $\B\in\Fq^{t_R\times n}$ and $\Hsub\in\Fqm^{(n-k-t_R)\times n}$. $\H_{\text{sub},i}$ the $i$-th row of $\Hsub$.}\label{fig:rank_figure}
\end{figure}

The relation between the error matrix $\E$, the matrix $\Hsub$, and the error supports for the Hamming metric, rank metric, and sum-rank metric are illustrated in Figures~\ref{fig:hamming_figure}, \ref{fig:rank_figure}, and \ref{fig:sum_rank_figure}, respectively. In particular, Figure~\ref{fig:sum_rank_figure} demonstrates the process of determining the sum-rank support $\SumRankSupp(\E)$ from the row spaces of the blocks $\H_{\CECode}^{(i)}$ for $i\inshots$.

\begin{figure}[ht!]
    \centering
    \begin{tikzpicture}[scale=.5]

    \draw[rounded corners, dashed] (-1, -1.5) rectangle (10, 5) {};
    \node[] () at (-0.5, 4.5) {$\E$};

    \draw[nonerrorblock] (0,0) rectangle (1,3);
    \draw[nonerrorblock] (1,0) rectangle (2,3);
    \draw[errorblock] (2,0) rectangle (3,3);
    \draw[nonerrorblock] (3,0) rectangle (5,3);
    \draw[errorblock] (5,0) rectangle (7,3);
    \draw[errorblock] (7,0) rectangle (8,3);
    \draw[nonerrorblock] (8,0) rectangle (9,3);

    \draw[-latex] (2.5,-0.5) -- (2.5,0);
    \draw[-latex] (6.0,-0.5) -- (6.0,0);
    \draw[-latex] (7.5,-0.5) -- (7.5,0);
    \node[] () at (5.0, -1.0) {blocks with rank errors};

    \node () at (0.5, 3.7) {$\E^{(1)}$};
    \node () at (4.5, 3.7) {$\cdots$};
    \node () at (8.8, 3.7) {$\E^{(\shots)}$};

    \node[] () at (10.7, 1.5) {$=$};

    \draw[rounded corners, dashed] (11.2, -1.5) rectangle (17.2, 5) {};
    \node[] () at (11.7, 4.5) {$\A$};

    \draw[errorblock] (12.5,0) rectangle (13.5,3);
    \draw[errorblock] (13.5,0) rectangle (14.0,3);
    \draw[errorblock] (14.0,0) rectangle (15.5,3);
    \node[] () at (12.8, 3.7) {$\A^{(1)}$};
    \node () at (14.4, 3.7) {$\cdots$};
    \node () at (15.8, 3.7) {$\A^{(\shots)}$};

    \node[] () at (17.8, 1.5) {$\cdot$};

    \draw[rounded corners, dashed] (18.3, -1.5) rectangle (27.8, 5) {};
    \node[] () at (18.8, 4.5) {$\B$};

    \draw[nonerrorblock] (18.8,0) rectangle (20.8,3);
    \draw[nonerrorblock] (20.8,0) rectangle (21.8,3);
    \draw[errorblock] (20.8, 2) rectangle (21.8, 3);
    \draw[nonerrorblock] (21.8,0) rectangle (23.8,3);
    \draw[nonerrorblock] (23.8,0) rectangle (25.8,3);
    \draw[errorblock] (23.8, 1.5) rectangle (25.8, 2);
    \draw[nonerrorblock] (25.8,0) rectangle (26.8,3);
    \draw[errorblock] (25.8, 0) rectangle (26.8, 1.5);

    \node () at (19.6, 3.7) {$\B^{(1)}$};
    \node () at (22.8, 3.7) {$\cdots$};
    \node () at (26.7, 3.7) {$\B^{(\shots)}$};

    \draw[-latex] (21.3, -0.5) -- (21.3, 0);
    \draw[-latex] (24.8, -0.5) -- (24.8,0);
    \draw[-latex] (26.3,-0.5) -- (26.3,0);
    \node[] () at (23.3, -1.0) {error blocks};

    \node[] () at (-1.5, -5) {$\Hce=$};
    \draw[errorblock] (0, -3.5) rectangle (2, -6.5);
    \draw[nonerrorblock] (2, -3.5) rectangle (3, -6.5);
    \draw[errorblock] (3, -3.5) rectangle (5, -6.5);
    \draw[nonerrorblock] (5, -3.5) rectangle (7, -6.5);
    \draw[errorblock] (7, -3.5) rectangle (9, -6.5);

    \node () at (1, -2.7) {$\Hce^{(1)}$};
    \node () at (4.5, -2.7) {$\cdots$};
    \node () at (8, -2.7) {$\Hce^{(\shots)}$};

    \draw[-latex] (2.5,-7.5) -- (2.5,-6.5);
    \draw[-latex] (6.0,-7.5) -- (6.0,-6.5);
    \node[] () at (4.5, -8.0) {all-zero blocks at position of full-rank errors};

    \node[] () at (13.5, -10) {$\SumRankSupp{(\E)} = \left(\mksum\limits_{j=1}^{n-k-\errWeight_{\Sigma R}} \RankSupp(\H_{\CECode,j}^{(1)})\right)^{\perp} \times\dots \times\left(\mksum\limits_{j=1}^{n-k-\errWeight_{\Sigma R}} \RankSupp(\H_{\CECode,j}^{(\intOrder)})\right)^{\perp}$};

\end{tikzpicture}
    \caption{Illustration of the error support for the sum-rank-metric case with $\E=\A\B\in\Fqm^{\intOrder \times n}$, $\A\in\Fqm^{\intOrder\times t_{\Sigma R}}$, $\B\in\Fq^{t_{\Sigma R}\times n}$ and $\Hsub\in\Fqm^{(n-k-t_{\Sigma R})\times n}$. $\A^{(i)}$ and $\B^{(i)}$ are the $i$-th block of $\A$ and $\B$ and $\H_{\CECode,j}^{(i)}$ the $j$-th row of $\Hsub^{(i)}$.}\label{fig:sum_rank_figure}
\end{figure}

\section{Conclusion}

In this paper, we consider a Metzner--Kapturowski-like decoding algorithm tailored for high-order interleaved sum-rank-metric codes. By leveraging the novel concept of an error code, we provided a fresh perspective on the decoding process. This approach not only enhances our understanding of the decoder's functionality but also offers new insights.

Our proposed algorithm demonstrates significant versatility, being applicable to any linear constituent code, including those that are unstructured or random. This general applicability positions our decoder as a robust tool for a wide range of coding scenarios. Furthermore, the computational complexity of our algorithm, which is on the order of $\oh{\max\{n^3, n^2s\}}$ operations over $\Fqm$, is independent of the code structure of the constituent code. This independence underscores the potential of our approach for practical implementations.

We also explored the success probability of our decoder, both within and beyond the unique decoding radius. Our analysis revealed that the decoder maintains a high success probability even for error weights exceeding the unique decoding radius. 

Our work not only extends the results of previous studies but also provides valuable insights for the design and security analysis of code-based cryptosystems based on interleaved sum-rank-metric codes.

The algorithm considered in this work is designed for interleaved codes in which the constituent codes are aligned vertically, also called vertically interleaved codes. From the error code perspective, this means that the error code's row support is restricted. In an alternative model, the codewords could be aligned horizontally (horizontal interleaving), resulting in the error code's column support being restricted. The adaptation of the considered algorithm to this horizontal interleaving model is not straightforward and remains an open problem.

Future research could explore further optimizations of the decoding algorithm and its application to other metrics.

\bibliography{references}

\begin{thebibliography}{10}
\providecommand{\url}[1]{#1}
\csname url@samestyle\endcsname
\providecommand{\newblock}{\relax}
\providecommand{\bibinfo}[2]{#2}
\providecommand{\BIBentrySTDinterwordspacing}{\spaceskip=0pt\relax}
\providecommand{\BIBentryALTinterwordstretchfactor}{4}
\providecommand{\BIBentryALTinterwordspacing}{\spaceskip=\fontdimen2\font plus
\BIBentryALTinterwordstretchfactor\fontdimen3\font minus \fontdimen4\font\relax}
\providecommand{\BIBforeignlanguage}[2]{{%
\expandafter\ifx\csname l@#1\endcsname\relax
\typeout{** WARNING: IEEEtran.bst: No hyphenation pattern has been}%
\typeout{** loaded for the language `#1'. Using the pattern for}%
\typeout{** the default language instead.}%
\else
\language=\csname l@#1\endcsname
\fi
#2}}
\providecommand{\BIBdecl}{\relax}
\BIBdecl

\bibitem{NISTreport2022}
G.~Alagic, D.~Apon, D.~Cooper, Q.~Dang, T.~Dang, J.~Kelsey, J.~Lichtinger, Y.-K. Liu, C.~Miller, D.~Moody, R.~Peralta, R.~Perlner, A.~Robinson, and D.~Smith-Tone, ``{Status Report on the Third Round of the {NIST} Post-Quantum Cryptography Standardization Process},'' 2022.

\bibitem{McEliece1978}
R.~J. McEliece, ``{A Public-Key Cryptosystem Based On Algebraic Coding Theory},'' \emph{The Deep Space Network Progress Report}, vol. 42-44, pp. 114--116, 1978.

\bibitem{elleuch2018public}
M.~Elleuch, A.~Wachter-Zeh, and A.~Zeh, ``{A Public-Key Cryptosystem from Interleaved Goppa Codes},'' \emph{arXiv preprint arXiv:1809.03024}, 2018.

\bibitem{holzbaur2019decoding}
L.~Holzbaur, H.~Liu, S.~Puchinger, and A.~Wachter-Zeh, ``{On Decoding and Applications of Interleaved Goppa Codes},'' in \emph{2019 IEEE International Symposium on Information Theory (ISIT)}, 2019, pp. 1887--1891.

\bibitem{renner2019interleavingloidreau}
J.~Renner, S.~Puchinger, and A.~Wachter-Zeh, ``{Interleaving Loidreau’s Rank-Metric Cryptosystem},'' in \emph{2019 XVI International Symposium "Problems of Redundancy in Information and Control Systems" (REDUNDANCY)}.\hskip 1em plus 0.5em minus 0.4em\relax IEEE, 2019, pp. 127--132.

\bibitem{krachkovsky1997decoding}
V.~Y. Krachkovsky and Y.~X. Lee, ``{Decoding for Iterative Reed--Solomon Coding Schemes},'' \emph{IEEE Transactions on Magnetics}, vol.~33, no.~5, pp. 2740--2742, 1997.

\bibitem{Loidreau_Overbeck_Interleaved_2006}
P.~Loidreau and R.~Overbeck, ``{Decoding Rank Errors Beyond the Error Correcting Capability},'' in \emph{International Workshop on Algebraic and Combinatorial Coding Theory (ACCT)}, Sep. 2006, pp. 186--190.

\bibitem{bartz2022fast}
H.~Bartz and S.~Puchinger, ``{Fast Decoding of Interleaved Linearized Reed–-Solomon Codes and Variants},'' \emph{submitted to: IEEE Transactions on Information Theory}, 2022, available at https://arxiv.org/abs/2201.01339.

\bibitem{metzner1990general}
J.~J. Metzner and E.~J. Kapturowski, ``{A General Decoding Technique Applicable to Replicated File Disagreement Location and Concatenated Code Decoding},'' \emph{IEEE Transactions on Information Theory}, vol.~36, no.~4, pp. 911--917, 1990.

\bibitem{renner2021decoding}
J.~Renner, S.~Puchinger, and A.~Wachter-Zeh, ``{Decoding High-Order Interleaved Rank-Metric Codes},'' in \emph{2021 IEEE International Symposium on Information Theory (ISIT)}.\hskip 1em plus 0.5em minus 0.4em\relax IEEE, 2021, pp. 19--24.

\bibitem{jerkovits2023highorder}
T.~Jerkovits, F.~H{\"o}rmann, and H.~Bartz, ``{On Decoding High-Order Interleaved Sum-Rank-Metric Codes},'' in \emph{Code-Based Cryptography}, J.-C. Deneuville, Ed.\hskip 1em plus 0.5em minus 0.4em\relax Cham: Springer Nature Switzerland, 2023, pp. 90--109.

\bibitem{hormannDistinguishingRecoveringGeneralized2023a}
\BIBentryALTinterwordspacing
F.~Hörmann, H.~Bartz, and A.-L. Horlemann, ``\BIBforeignlanguage{en}{Distinguishing and {Recovering} {Generalized} {Linearized} {Reed}–{Solomon} {Codes}},'' in \emph{\BIBforeignlanguage{en}{Code-{Based} {Cryptography}}}, J.-C. Deneuville, Ed.\hskip 1em plus 0.5em minus 0.4em\relax Cham: Springer Nature Switzerland, 2023, vol. 13839, pp. 1--20, series Title: Lecture Notes in Computer Science. [Online]. Available: \url{https://link.springer.com/10.1007/978-3-031-29689-5_1}
\BIBentrySTDinterwordspacing

\bibitem{puchingerGenericDecodingSumRank2022}
S.~Puchinger, J.~Renner, and J.~Rosenkilde, ``Generic {Decoding} in the {Sum}-{Rank} {Metric},'' \emph{IEEE Transactions on Information Theory}, vol.~68, no.~8, 2022.

\bibitem{matsaglia_equalities_1974}
\BIBentryALTinterwordspacing
G.~Matsaglia and G.~P.~H.~Styan, ``\BIBforeignlanguage{en}{Equalities and {Inequalities} for {Ranks} of {Matrices} $^{\textrm{†}}$},'' \emph{\BIBforeignlanguage{en}{Linear and Multilinear Algebra}}, vol.~2, no.~3, pp. 269--292, Jan. 1974. [Online]. Available: \url{http://www.tandfonline.com/doi/abs/10.1080/03081087408817070}
\BIBentrySTDinterwordspacing

\bibitem{puchinger2019decoding}
S.~Puchinger, J.~Renner, and A.~Wachter-Zeh, ``{Decoding High-Order Interleaved Rank-Metric Codes},'' \emph{arXiv preprint arXiv:1904.08774}, 2019.

\bibitem{horlemann2021information}
A.-L. Horlemann, S.~Puchinger, J.~Renner, T.~Schamberger, and A.~Wachter-Zeh, ``{Information-Set Decoding with Hints},'' in \emph{Code-Based Cryptography Workshop}.\hskip 1em plus 0.5em minus 0.4em\relax Springer, 2021, pp. 60--83.

\bibitem{Couveignes2009}
J.-M. Couveignes and R.~Lercier, ``{Elliptic Periods for Finite Fields},'' \emph{Finite Fields and Their Applications}, vol.~15, no.~1, pp. 1--22, 2009.

\bibitem{puchinger2020generic}
S.~Puchinger, J.~Renner, and J.~Rosenkilde, ``{Generic Decoding in the Sum-Rank Metric},'' in \emph{2020 IEEE International Symposium on Information Theory (ISIT)}.\hskip 1em plus 0.5em minus 0.4em\relax IEEE, 2020, pp. 54--59.

\bibitem{lidl_finite_1996}
R.~Lidl and H.~Niederreiter, \emph{Finite {Fields}}, ser. Encyclopedia of {Mathematics} and its {Applications}.\hskip 1em plus 0.5em minus 0.4em\relax Cambridge University Press, Oct. 1996, published: Hardcover.

\bibitem{sagemath}
{The Sage Developers}, \emph{{S}ageMath, the {S}age {M}athematics {S}oftware {S}ystem ({V}ersion 9.8)}, 2023, {\tt https://www.sagemath.org}.

\end{thebibliography}
\bibliographystyle{IEEEtran}

\end{document}